\documentclass[sigconf,authorversion,nonacm]{acmart}

\AtBeginDocument{%
  \providecommand\BibTeX{{%
    \normalfont B\kern-0.5em{\scshape i\kern-0.25em b}\kern-0.8em\TeX}}}

\usepackage{graphicx} 
\usepackage{url}
\usepackage{mathtools}
\usepackage{physics}
\usepackage{xcolor}
\usepackage{tikz}
\usepackage{subfigure}
\usepackage{algorithm}
\usepackage[toc,page]{appendix}

\DeclareMathOperator*{\argmin}{arg\,min}

\newtheorem{definition}{Definition}
\newtheorem{problem}{Problem}
\newtheorem{remark}{Remark}

\newcommand{\ncf}{\delta}

\newtheorem{proposition}{Proposition}

\newcommand{\camera}[1]{#1}

\newcommand\copyrightnotice[1]{
    \begin{tikzpicture}[remember picture,overlay]
    \node[anchor=south,yshift=10pt] at (current page.south) {\fbox{\parbox{\dimexpr\textwidth-\fboxsep-\fboxrule\relax}{#1}}};
    \end{tikzpicture}
}

\begin{document}

\title{Conformal Quantitative Predictive Monitoring of STL Requirements for Stochastic Processes}

\author{Francesca Cairoli}
\affiliation{%
  \institution{University of Trieste}
  \city{Trieste}
  \country{Italy}}
\email{francesca.cairoli@units.it}

\author{Nicola Paoletti}
\affiliation{%
  \institution{King's College London}
  \city{London}
  \country{United Kingdom}}
\email{nicola.paoletti@kcl.ac.uk}

\author{Luca Bortolussi}
\affiliation{%
  \institution{University of Trieste}
  \city{Trieste}
  \country{Italy}}
\email{lbortolussi@units.it}

\begin{abstract}

We consider the problem of predictive monitoring (PM), i.e., predicting at runtime the satisfaction of a desired property from the current system's state. Due to its relevance for runtime safety assurance and online control, PM methods need to be efficient to enable timely interventions against predicted violations, while providing correctness guarantees. 
We introduce \textit{quantitative predictive monitoring (QPM)}, the first PM method to support stochastic processes and rich specifications given in Signal Temporal Logic (STL). Unlike most of the existing PM techniques that predict whether or not some property $\phi$ is satisfied, QPM provides a quantitative measure of satisfaction by predicting the quantitative (aka robust) STL semantics of $\phi$. QPM derives prediction intervals that are highly efficient to compute and with probabilistic guarantees, in that the intervals cover with arbitrary probability the STL robustness values relative to the stochastic evolution of the system.  To do so, we take a machine-learning approach and leverage recent advances in conformal inference for quantile regression, thereby avoiding expensive Monte Carlo simulations at runtime to estimate the intervals. 
We also show how our monitors can be combined in a compositional manner to handle composite formulas, without retraining the predictors or sacrificing the guarantees. 
We demonstrate the effectiveness and scalability of QPM over a benchmark of four discrete-time stochastic processes with varying degrees of complexity. 
\end{abstract}


\acmConference[]{}{}{}
\setcopyright{none}

\maketitle

\copyrightnotice{\textbf{\copyright ACM 2023. This is the author's version of the work. It is posted here for your personal use. Not for redistribution. The definitive Version of Record was published in Proceedings of the 26th ACM International Conference on Hybrid Systems: Computation and Control, \url{http://dx.doi.org/10.1145/3575870.3587113}.}}

\section{Introduction}\label{sec:introduction}

Predictive monitoring (PM) ~\cite{bortolussi2019neural,bortolussi2021neural,cairoli2021neural} is the problem of predicting at runtime the satisfaction of a certain requirement from the current system's state. Unlike traditional monitoring~\cite{bartocci2018specification}, PM has the potential to detect failures before they occur, thereby enabling preemptive countermeasures, such as switching to a fail-safe mode~\cite{phan2020neural}.  To enable effective deployment at runtime, PM methods need to be efficient and respond quickly, so that any system failure can be prevented in time. 

Performing model checking at run-time would provide a precise solution to the PM problem (precise up to the accuracy of the system's model), but such a solution is computationally expensive in general, unless the model is fully deterministic. In particular, when the system is stochastic, a statistical model checking solution~\cite{younes2006numerical} would require simulating at runtime a typically large number of Monte-Carlo trajectories to achieve desired error levels\footnote{Numerical/symbolic probabilistic model checking techniques typically are at least as expensive as statistical methods and can be applied only to a restricted class of models.}. 

For this reason, a number of approximate PM techniques based on machine learning have been recently proposed (see e.g.~\cite{qin2019predictive,yel2020assured,chou2020predictive,ma2021predictive}), including the so-called Neural Predictive Monitoring (NPM) method~\cite{bortolussi2019neural,bortolussi2021neural,cairoli2021neural}. In NPM in particular, the predictive monitor is a neural network classifier trained using data generated through a model checker to predict for any system state whether or not the state satisfies some reachability property. To improve the reliability of the reachability predictions, NPM relies on conformal prediction~\cite{balasubramanian2014conformal,vovk2005algorithmic} to produce prediction regions with guaranteed coverage and derive  uncertainty measures used to infer whether a particular prediction can be trusted. 

In this paper, we introduce a predictive monitoring method for stochastic processes and Signal Temporal Logic (STL)~\cite{donze2010robust} specifications. We call it \textit{quantitative predictive monitoring (QPM)} as it represents, to the best of our knowledge, the first method that can predict the quantitative semantics $R_{\phi}$ of an STL formula $\phi$ (aka the STL robustness) for a general class of stochastic systems and with probabilistic validity guarantees. QPM is inspired by NPM but it addresses two significant limitations of the latter, which supports only Boolean reachability specifications (as opposed to the full spectrum of STL properties and their quantitative interpretation), and cannot adequately deal with stochastic dynamics.

By supporting arbitrary STL specifications, with QPM we gain {expressiveness} in the type of requirements that can be actually monitored and thus in the type of violations that can be detected. In particular, by predicting STL robustness values, QPM provides key quantitative information on the degree of property satisfaction, unlike existing predictive monitoring approaches that can only predict Boolean satisfaction. STL robustness predictions can be meaningfully used to determine the extent of any corrective actions and enable efficient online model predictive control for STL~\cite{raman2014model}: for instance, depending on the specification $\phi$, if QPM predicts a high value of $R_{\phi}$, then little or no intervention might be needed, while a low $R_{\phi}$ value might require a more substantial intervention to steer the system back to safety. 

A major challenge when dealing with stochastic processes is that every state induces a distribution of robustness values (conditional on the state), relative to the future, time-bounded, stochastic evolution of the system. In general, such distribution is analytically intractable\footnote{With simple stochastic models and simple properties, it may become tractable but still impractical to analyze at runtime.} and an accurate empirical estimate of such conditional distribution can be very expensive to obtain, potentially requiring a high number of Monte-Carlo simulations. 

Our approach overcomes this computational bottleneck by deriving monitors able to directly predict some relevant quantiles of the conditional STL robustness distribution. Such quantiles have a two-fold purpose. First, they provide a \textit{measure of risk}~\cite{majumdar2020should,lindemann2022temporal}: for instance, if the 10\%-quantile of $R_{\phi}$ is zero, then ``only'' 10\% of the system's future trajectories will violate $\phi$ (i.e., lead to a negative $R_{\phi}$), which, depending on the application, can be interpreted as a low-risk scenario. Second, and most crucially, it allows us to derive \textit{prediction intervals} that cover a certain mass of probability. For small $\alpha\in(0,0.5)$, if the $(\alpha,1-\alpha)$-quantile interval for $R_{\phi}$ is entirely above (below) zero, then we can be fairly confident that the system will evolve into satisfying (violating) $\phi$ with high probability. On the other hand, intervals straddling the zero denote states whose future trajectories may or may not violate the property, which we, therefore, label as uncertain.   

To predict the quantiles in a reliable manner, our approach builds on \textit{Conformalized Quantile Regression (CQR)}~\cite{romano2019conformalized}, a recent conformal inference technique that produces statistically valid prediction intervals on top of quantile regression models. Crucially, this method provides us with probabilistic guarantees, in that, for an arbitrary level $\alpha$, the resulting interval for $R_{\phi}$ is guaranteed to include the true STL robustness value with probability at least $1-2\alpha$. Such guarantees are akin to those of statistical model checking, with the key difference that CQR prediction intervals are extremely fast to compute and thus, can be used at runtime.

By default, our QPM approach derives a monitor for a fixed choice of STL property $\phi$. On one hand, this is advantageous in that the monitor is tailored, and hence, highly accurate at predicting $R_{\phi}$ but on the other hand, it lacks flexibility. To this purpose, we further show how monitors trained for different properties $\phi_1$ and $\phi_2$ can be combined in a modular way to monitor Boolean combinations of $\phi_1$ and $\phi_2$, without retraining the underlying models and preserving the desired probabilistic guarantees,

In summary, our contributions are:
\begin{itemize}
    \item[-] We introduce \textit{quantitative predictive monitoring (QPM)}, the first predictive monitoring method to support stochastic processes and the quantitative semantics of arbitrary STL specifications. QPM builds on recent conformal inference techniques to provide prediction intervals for STL robustness with probabilistic guarantees.
    \item[-] We demonstrate the compositionality of our approach by showing that composite formulas can be monitored by combining the QPM monitors of the subformulas.
    \item[-] We evaluate the effectiveness and scalability of QPM on several case studies of stochastic processes. 
\end{itemize}

\section{Problem Statement}\label{sec:background}
We illustrate the predictive monitoring problem we target with QPM, after introducing background about stochastic processes and Signal Temporal Logic.

\subsection{Stochastic Processes}

The systems we consider can be modeled as stochastic processes. A stochastic process  is defined as a collection of random  variables indexed by some index set $T$. These random variables are defined on a common probability space $(\Omega,\mathcal{F},\mathbb{P})$, where $\Omega$ is the sample space, $\mathcal{F}$ is the $\sigma$-algebra and $\mathbb{P}$ is the probability measure. We can denote the stochastic process as $\{ {\mathbf{S}(t,\omega), t\in T\}}$. A random variable $\mathbf{S}(t,\omega)$ in the collection is thus a function of two variables $t\in T$ and $\omega\in\Omega$. In our application, the index set is countable and represents \emph{discrete time} $T=\{0,1,\ldots\}$. Each random variable in the collection takes values in a space $S\subseteq\mathbb{R}^n$, the \emph{state space} of dimension $n$, that should be measurable. A discrete-time step makes the stochastic process move from index $i$ to index $i+1$. Given a stochastic process ${\displaystyle \{\mathbf{S}(t,\omega ):t\in T\}}$, then for any point $\omega \in \Omega$, the mapping
${\displaystyle \mathbf{S}(\cdot ,\omega ):T\rightarrow S,}$
is called a \emph{realization}, or a \emph{sample trajectory} of the stochastic process ${\displaystyle \{\mathbf{S}(t,\omega ):t\in T\}}$. 
\camera{We assume that the dynamics of the system is Markovian. This assumption is not strict as most systems of interest -- Markov chains, stochastic hybrid systems (without non-determinism), and stochastic difference equations -- are Markovian or can be made so by augmenting the state space.}

\subsection{Signal Temporal Logic}\label{subsec:stl}
System requirements can be expressed via Signal Temporal Logic (STL)~\cite{maler2004monitoring,donze2010robust}, which
enables the specification of properties of dense-time, real-valued signals, and the automatic generation of monitors for
testing properties on individual trajectories. The rationale of STL is to transform real-valued signals into Boolean ones, using formulae built on the following \emph{STL syntax}: $\displaystyle \phi := true\ |\ \mu\ |\ \neg\phi\ |\  \phi\land\phi\ |\ \phi\ U_I\phi,$
where $I\subseteq \mathbb{T}$ is a temporal interval, either bounded, $I = [a,b]$, or unbounded, $I = [a,+\infty)$, for any $0\le a < b$. Atomic propositions $\mu$ are (non-linear) inequalities on the states of a signal $\Vec{s}$ at a time $t$, $\mu = (g(\Vec{s}(t))>0 )$, where $g:S\to\mathbb{R}$ and $\Vec{s}(t)$ is a state in $S$. 
From this essential syntax, it is easy to define other operators, used to abbreviate the syntax in an STL formula: $false :=\neg true$, $\phi \lor \psi := \neg(\neg\phi\land\neg\psi)$, $F_I:= true\ U_I\phi$ and $G_I:= \neg F_I\neg\phi$. Monitoring the satisfaction of a formula is done recursively by leveraging the tree structure of the STL formula. The satisfaction relation is defined as follows.

\vspace{-0.25cm}

\begin{align*}
&(\Vec{s}, t) \models \mu &\Leftrightarrow &\hspace{2ex} g(\Vec{s}(t))>0 \vspace{0.02in}\\
&(\Vec{s}, t) \models \neg \phi &\Leftrightarrow &\hspace{2ex} \neg((\Vec{s}, t) \models \phi)\vspace{0.02in}\\
&(\Vec{s},t) \models \phi_1 \wedge \phi_2  &\Leftrightarrow&\hspace{2ex} (\Vec{s},t) \models \phi_1 \wedge (\Vec{s},t) \models \phi_2 \vspace{0.02in}\\
&(\Vec{s},t) \models \phi_1 U_{a,b} \phi_2 &\Leftrightarrow&\hspace{2ex} \exists\ t'\in [t + a,t+b]\ \text{s.t.}\ (\Vec{s},t') \models \phi_2 \wedge \\
&&&\hspace{6em}\forall\; t''\in [t,t'],\ (\Vec{s},t'') \models \phi_1 
\end{align*}

\paragraph{Quantitative semantics.} The robustness of a trajectory quantifies the level of satisfaction w.r.t. $\phi$. Positive robustness means that the property is satisfied, whereas negative robustness means that the property is violated. Robustness is denoted as a function $R_{\phi}: S^H\times T \rightarrow \mathbb{R}$ that maps a given signal $\Vec{s}$ of length $H$, a formula $\phi$ and a time $t$ to some real value, $R_\phi(\Vec{s},t)\in\mathbb{R}$. It measures the maximum perturbation that can be applied to the signal without changing its truth value w.r.t. $\phi$. 

Similarly to the Boolean semantics, the quantitative semantics of a formula $\phi$ over a signal $\Vec{s}$ is defined recursively over the tree structure of the STL formula, as described below:
\begin{align*}
    & R_\mu (\Vec{s},t) &=& \ g(\Vec{s}(t))\\
    & R_{\neg\phi}(\Vec{s},t) &=& -R_{\phi}(\Vec{s},t)\\
    & R_{\phi_1\land\phi_2}(\Vec{s},t) &=& \min (R_{\phi_1}(\Vec{s},t),R_{\phi_2}(\Vec{s},t))\\
    & R_{\phi_1 U_{[a,b]}\phi_2}(\Vec{s},t) &=&
    \underset{\scriptscriptstyle t'\in [t+a,t+b]}{\sup}\Big(\min\big(R_{\phi_2}(\Vec{s},t'), \underset{\scriptscriptstyle t''\in [t,t']}{\inf}R_{\phi_1}(\Vec{s},t'')\big)\Big).
\end{align*}
The sign of $R_\phi$ indicates the Boolean satisfaction of signal $\Vec{s}$, and in particular, we have
\begin{itemize}
    \item $R_\phi(\Vec{s},t)>0\Rightarrow (\Vec{s},t)\models\phi$;
    \item $R_\phi(\Vec{s},t)<0\Rightarrow (\Vec{s},t)\not\models\phi$.
\end{itemize}

Given a stochastic process $\mathbf{S} = \{\mathbf{S}(t,\omega),t\in T\}$, an STL requirement $\phi$ and a state $s_0\in S$, the robustness over future evolutions of the system starting from $s_0$ is stochastically distributed according to the conditional distribution $$\mathbb{P}\left(R_\phi(\Vec{s},\camera{0}) \mid \Vec{s}(0)=s_0\right),$$ where $\Vec{s}$ is a random signal given by the sequence of random variables $(\mathbf{S}(0,\cdot), \mathbf{S}(1,\cdot), \ldots, \mathbf{S}(H-1,\cdot))$. For time $k$, 
$\Vec{s}(k)=\mathbf{S}(k,\cdot)$ denotes the random variable corresponding to the state at time $k$ in $\Vec{s}$. This conditional distribution captures the distribution of the STL robustness values for trajectories of length $H$ starting in $s_0$.

\noindent We now formulate the quantitative predictive monitoring problem: from any state $s_*$ of the stochastic process, we aim to construct a prediction interval guaranteed to include, with arbitrary probability, the true STL robustness of any (unknown) stochastic trajectory starting at $s$. A formal statement of the problem is given below. 

\begin{problem}[Quantitative Predictive Monitoring]\label{prbl:qpm}
Given a discrete-time stochastic process $\mathbf{S} = \{\mathbf{S}(t,\omega),t\in T\}$ over a state space $S$,  temporal horizon $H$, 
a significance level $\alpha\in [0,1]$ and an STL formula $\phi$, derive a monitoring function $I$ that maps any state $s_* \sim \mathbf{S}(\cdot,\cdot)$ into an interval $I(s_*)$ such that
$$\mathbb{P}\left(R_\phi(\Vec{s},0) \in I(s_*)|\Vec{s}(0)=s_*\right) \geq 1-\alpha.$$



\end{problem}

We frame Problem~\ref{prbl:qpm} as a conditional quantile regression problem. This boils down to learning for a generic state $s_*$ an upper and a lower quantile of the random variable $R_\phi(\Vec{s},0)$ induced by $\Vec{s}$ conditioned on $\Vec{s}(0)=s_*$. We then use these two quantiles to build the output of function $I$ in $s_*$. To ensure that such an interval is well-calibrated, meaning that the probabilistic guarantees are satisfied theoretically and empirically, we resort to the framework of conformal prediction. These machine-learning techniques are introduced in the next section.

\section{Background on Conformal Prediction}\label{subsec:cp}
Consider a generic supervised learning framework where $X$ denotes the input space, $Y$ the target space, and $Z = X\times Y$. Let $\mathcal{Z}$ be the data-generating distribution, i.e., the distribution of the points $(x,y)\in Z$.  
We assume that the target $y$ of a point $(x, y)\in Z$ is the result of the application of
a function $f^*:X\to Y$, typically unknown or very expensive to evaluate. The goal of a supervised learning algorithm is to find a function $f: X\rightarrow Y$ that, from a finite set of observations, learns to behave as similarly as possible to $f^*$ over the entire input space. 
For an input $x\in X$, 
we denote with $t$ the true target value of $x$ and with $\hat{y}$ the prediction by $f$, i.e. $\hat{y} = f(x)$. 
Test inputs, whose unknown true target values we aim to predict, are denoted by $x_*$. 
For the sake of clarity, we start by showing conformal prediction approaches for deterministic predictors and then move to present {Conformalized Quantile Regression (CQR)}~\cite{romano2019conformalized}, an approach to handle the stochastic case. 

\emph{Conformal predictions} (CP) associate measures of reliability with any traditional supervised learning problem, either regression or classification~\cite{balasubramanian2014conformal,vovk2005algorithmic}. CP enriches point-wise predictions with \emph{prediction regions with guaranteed validity}.

\begin{definition}[Prediction region]
For significance level $\alpha \in (0,1)$ and test input $x_*$, the $\alpha$-prediction region for $x_*$, $\Gamma_*^{\alpha}\subseteq Y$, is a set of target values s.t.
\begin{equation}\label{eq:pred_r}
    \mathbb{P}_{(x_*,y_*)\sim \mathcal{Z}}(y_* \in \Gamma_*^{\alpha}) \ge 1 - \alpha.
\end{equation}
\end{definition}

The idea of CP is to construct the prediction region by ``inverting'' a suitable hypothesis test: given a test point $x_*$ and a tentative target value $y'$, we \textit{exclude} $y'$ from the prediction region only if it is unlikely that $y'$ is the true value for $x_*$. The test statistic is given by a so-called \textit{nonconformity function (NCF)} $\ncf:Z \rightarrow \mathbb{R}$, which, given a predictor $f$ and a point $z=(x,y)$, measures the deviation between the true value $y$ and the corresponding prediction $f(x)$.  In this sense, $\ncf$ can be viewed as a generalized residual function. In other words, CP builds the prediction region $\Gamma_*^{\alpha}$ for a test point $x_*$ by excluding all targets $t'$ whose NCF values are unlikely to follow the NCF distribution of the true targets:
\begin{equation}\label{eq:cp_predr}
\Gamma_*^{\alpha} = \left\{y' \in Y \mid  Pr_{(x,y)\sim \mathcal{Z}}\left(\ncf(x_*,y') \geq \ncf(x,y)\right) > \alpha\right\}.
\end{equation}
This prediction region is guaranteed to contain the true (unknown) value $y_*$ with confidence $1-\alpha$.
The probability term in Eq.~\eqref{eq:cp_predr} is often called the p-value. 
From a practical viewpoint, the NCF distribution $Pr_{(x,y)\sim \mathcal{Z}}(\ncf(x,y))$ cannot be derived in an analytical form, and thus we use an empirical approximation derived using a sample $Z_c$ of $\mathcal{Z}$.  This  approach is called \textit{inductive CP}~\cite{papadopoulos2008inductive} and $Z_c$ is referred to as \textit{calibration set}. 
 CP's theoretical guarantees hold under the \textit{exchangeability} assumption (a ``relaxed'' version of i.i.d.) by which the joint probability of any sample of $\mathcal{Z}$ is invariant to permutations of the sampled points.

\paragraph{Validity and Efficiency.} 
CP performance is measured via two quantities: 1) \emph{validity} (or \emph{coverage}), i.e. the empirical error rate observed on a test sample, which should be as close as possible to the significance level $\alpha$, and 2) \emph{efficiency}, i.e. the size of the prediction regions, which should be small. CP-based prediction regions are automatically valid, whereas the efficiency depends on the chosen nonconformity function and on the accuracy of the underlying model.

\subsection{Conformal Prediction for Regression}\label{sec:cp_regr}

In regression problems, we have a continuous target space $Y\subseteq\mathbb{R}^n$. 
The inductive CP algorithm is divided into an offline phase, executed only once, and an online phase, executed for every test point $x_*$. In the offline phase (steps 1--3 below), we train the classifier $f$ and construct the calibration distribution, i.e., the empirical approximation of the NCF distribution. In the online phase (steps 4--5), we derive the prediction region for $x_*$ using the computed regressor and distribution.
\begin{enumerate}
    \item Draw sample $Z'$ of $\mathcal{Z}$. Split $Z'$ into training set $Z_t$ and calibration set $Z_c$.
    \item Train regressor $f$ using $Z_t$. Use $f$ to define an NCF $\delta$.
    \item Construct the calibration distribution by computing, for each $z_i \in Z_c$, the NCF score $\beta_i = \delta(z_i)$.
    \item Identify the critical value $\beta_{(\alpha)}$ of the calibration distribution, i.e. its empirical $(1-\alpha)$-quantile, or the $\lfloor \alpha \cdot (|Z_c|+1)\rfloor$-th largest calibration score.
\item Return the prediction region
\begin{equation}\label{eq:pred_reg}
   \Gamma_*^{\alpha} = f(x_*) \pm \beta_{(\alpha)}.
\end{equation}

\end{enumerate}

Notice that such  prediction intervals have the same width ($\beta_{(\alpha)}$) for all inputs. 
A natural NCF in regression is the norm of the difference between the real and the predicted target value, i.e., $\ncf(x) = ||y - f(x)||$. 

\paragraph{Predictive uncertainty.} A CP-based prediction region provides a set of plausible predictions with statistical guarantees, and as such, also captures the uncertainty about the prediction. 
The size of the prediction region is determined by the chosen significance level $\alpha$. 
Specifically, from Eq.~\eqref{eq:pred_reg} we can see that, for levels  $\alpha_1\ge\alpha_2$, the corresponding prediction regions are such that $\Gamma^{\alpha_1}\subseteq \Gamma^{\alpha_2}$, as a smaller $\alpha$ yields a larger critical value $\beta_{(\alpha)}$. 

\subsection{Conformalized Quantile Regression}\label{subsec:cqr}

We now switch to the stochastic setting. Let us consider a probabilistic function mapping an input $x\in X$ into a distribution over the target space $Y$. 

\paragraph{Quantile Regression.} The aim of conditional Quantile Regression (QR) is to estimate a given quantile of such a distribution over $Y$ conditional on an input $x\in X$. Let $F(y'|x=x') := \mathbb{P}(y\le y'|x=x')$ be the conditional distribution function of $y$ given $x$. Then, the $\alpha$-th conditional quantile function is defined as
\begin{equation}\label{eq:cond_quantile_fnc}
    q_\alpha := \inf \{y'\in\mathbb{R} \mid F(y'|x=x')\ge\alpha\}.
\end{equation}

Given a significance level $\alpha$, we consider lower and upper quantiles w.r.t.\ $\alpha_{lo} =\alpha /2$ and $\alpha_{hi}=1-\alpha /2$, respectively. We define the desired prediction interval as $PI(x) := [q_{\alpha_{lo}}(x), q_{\alpha_{hi}}(x)]$.
By construction, this interval satisfies 
\begin{equation}\label{eq:pi_validity}
\mathbb{P}(y\in PI(x')|x=x')\ge 1-\alpha.
\end{equation}
Since the prediction interval is conditional on the input, the length of the interval is not fixed in general and changes at different values of $x$.
QR infers such prediction interval from the data. In particular, estimating the quantiles can be expressed as approximating the quantile function and thus it can be framed as an optimization problem. In a nutshell, the idea is to propose a parametric function $f(x;\theta)$ as a candidate approximator for $q_\alpha (x)$ and then optimize over its parameters $\theta$ so that it closely resembles the quantile function. The optimization problem can be formally expressed as finding the optimal parameters $\hat{\theta}$ such that
$$\hat{\theta} = \argmin_\theta \left[\frac{1}{n}\sum_i \mathcal{L}_{\alpha}(Y_i,f(X_i;\theta))+\mathcal{G}(\theta)\right],$$ where $\mathcal{L}_\alpha$ is the check or pinball loss, defined as 
\begin{equation}\label{eq:loss}
\mathcal{L}_\alpha(y,\hat{y}) = \alpha\max(y-\hat{y},0)+(1-\alpha)\max(\hat{y}-y,0),
\end{equation}
and $\mathcal{G}$ is an optional regularization term.
In this work, we use deep neural networks (NN) as candidate parametric approximators $f(\cdot,\theta)$ of the quantile function and train them using Eq.~\eqref{eq:loss} as loss function. Once trained, the output of $f(x;\hat{\theta})$ is a prediction $\hat{q}_\alpha(x)$ of the conditional $\alpha$-th quantile.
In general, each quantile requires the training of a different neural network. However, one could also train a single multi-output NN that learns to approximate multiple quantile functions at the same time. The multi-output objective function is obtained by averaging over the respective losses. In this work, we decide to simultaneously learn $q_{\alpha_{lo}}$, $q_{0.5}$ (the median) and $q_{\alpha_{hi}}$, where for a choice of $\alpha\in(0,0.5)$, $\alpha_{lo}=\alpha$ and $\alpha_{hi}=1-\alpha$. The loss then becomes 
$$\mathcal{L}_\alpha(y,\hat{y}) = \tfrac{1}{3}\cdot (\mathcal{L}_{\alpha_{lo}}(y,\hat{y})+\mathcal{L}_{0.5}(y,\hat{y})+\mathcal{L}_{\alpha_{hi}}(y,\hat{y})).$$
In general, a single NN could simultaneously learn even a larger number of quantiles.

\paragraph{Conformalized Quantile Regression}

The goal of Conformalized Quantile Regression (CQR) is to adjust the QR prediction interval so that it is guaranteed to contain the $(1-\alpha)$ mass of probability, i.e. to satisfy~\eqref{eq:pi_validity}.
As for CP, we divide the dataset $Z'$ in a training set $Z_t$ and a calibration set $Z_c$. We train the QR $f(\cdot;\theta)$ over $Z_t$ and on $Z_c$ we compute the nonconformity scores as
\begin{equation}\label{eq:CQR_cal_set}
E_i := \max \{\hat{q}_{\alpha_{lo}}(x_i)-y_i, y_i-\hat{q}_{\alpha_{hi}}(x_i)\mid  (x_i,y_i)\in Z_c\}.
\end{equation}
In our notation, $\hat{q}_{\alpha_{lo}}(x)$ and $\hat{q}_{\alpha_{lo}}(x)$ denotes the two outputs of $f(x;\hat{\theta})$. 
The conformalized prediction interval is thus defined as
$$CPI(x_*) = [\hat{q}_{\alpha_{lo}}(x_*)-\tau, \hat{q}_{\alpha_{hi}}(x_*)+\tau],$$ where $\tau$ is the $ (1-\alpha)(1+1/|Z_c|)$-th empirical quantile of $\{E_i:i\in Z_c\}$.

In the following, we will abbreviate with \textit{PI} a (non-calibrated) QR prediction interval and with \textit{CPI} a (calibrated) conformalized prediction interval.

\begin{remark}
This nonconformity function, and thus $\tau$, can be negative and thus the conformalized prediction interval can be tighter than the original prediction interval.
This means that the CPI can be more efficient than the PI, where the efficiency is the average width of the prediction intervals over a test set.
The CPI has guaranteed coverage (the PI does not), meaning $\mathbb{P}_{(x_*,y_*)\sim \mathcal{Z}}(y_*\in CPI(x_*))\ge 1-\alpha$.
\end{remark} 

\begin{remark}
In principle, one could use traditional CP for regression (see Section~\ref{sec:cp_regr}) to obtain valid prediction intervals for the stochastic case. The main advantage of CQR is that it produces CPIs that are adaptive to heteroscedasticity, i.e., they account for the fact that the variability in the output may be affected by the value of the input. On the other hand, intervals produced by CP for regression have fixed sizes and hence, do not account for heteroscedasticity. Moreover, the PI of CP for regression would be relative to the conditional mean of the STL robustness, not to the conditional quantile range that we are interested in. 
\end{remark}

\section{Quantitative Predictive Monitoring}\label{sec:problem}
We present a method to solve Problem~\ref{prbl:qpm}\footnote{When the context is clear, we'll use the term ``Quantitative Predictive Monitoring'' to refer to both the problem and the solution method.}. For a given a discrete-time stochastic process $\mathbf{S} = \{\mathbf{S}(t,\omega),t\in T\}$ over state space $S$ and a state $s \sim \mathbf{S}(k,\cdot)$ at time $k \in T$, the stochastic evolution (bounded by horizon $H$) of the system starting at $s$ can be described by the conditional distribution 
$$\mathbb{P}(\Vec{s} \mid \Vec{s}(k) = s),$$
where $\Vec{s}=(\Vec{s}(k),\ldots, \Vec{s}(k+H))\in S^H$ is the random trajectory of length $H$ starting at time $t_k$,  $\Vec{s}(i)=\mathbf{S}(i,\cdot)$ for any $i\in T$. 

The quantitative STL semantics inherits the stochasticity from the dynamics of the system. For an STL property $\phi$, we denote with $\mathcal{R}_\phi(s)$ the random variable denoting the STL robustness value relative to $\phi$ of trajectories starting from $s\in S$ at time $t_k$, i.e.,
$$\mathcal{R}_\phi(s) \sim \mathbb{P}(R_\phi(\Vec{s},0) \mid \Vec{s}(k)=s).$$

Our conformal solution to the QPM aims at finding, for each state $s\in S$ of the system, a prediction region that covers a certain probability mass of the STL robustness distribution $\mathcal{R}_\phi(s)$, see Problem~\ref{prbl:qpm}. 
In simpler terms, we aim at monitoring how safe the system is relative to the unknown and stochastic future evolution from its current state $s$. In this way, one can intervene preemptively on the system in order to prevent any failures. 
However, the distribution of $\mathcal{R}_\phi(s)$ is 
impossible to compute exactly and in an efficient manner. Thus, we resort to Conformalized Quantile Regression (CQR), introduced in Section~\ref{subsec:cqr}, to compute a prediction interval that, for each state $s$, is guaranteed to cover a desired level $(1-\alpha)$ of the probability mass for the conditional distribution of robustness values $\mathcal{R}_\phi(s)$.

In short, our solution consists of four steps, detailed below: \textit{dataset generation}, \textit{QR training}, \textit{residuals computation}, and \textit{inference}. Note that only the last step, which is by far the quickest, is performed online, the others are performed offline and hence, do not affect runtime performance.

         \begin{figure}
              \centering
              \includegraphics[width=\columnwidth]{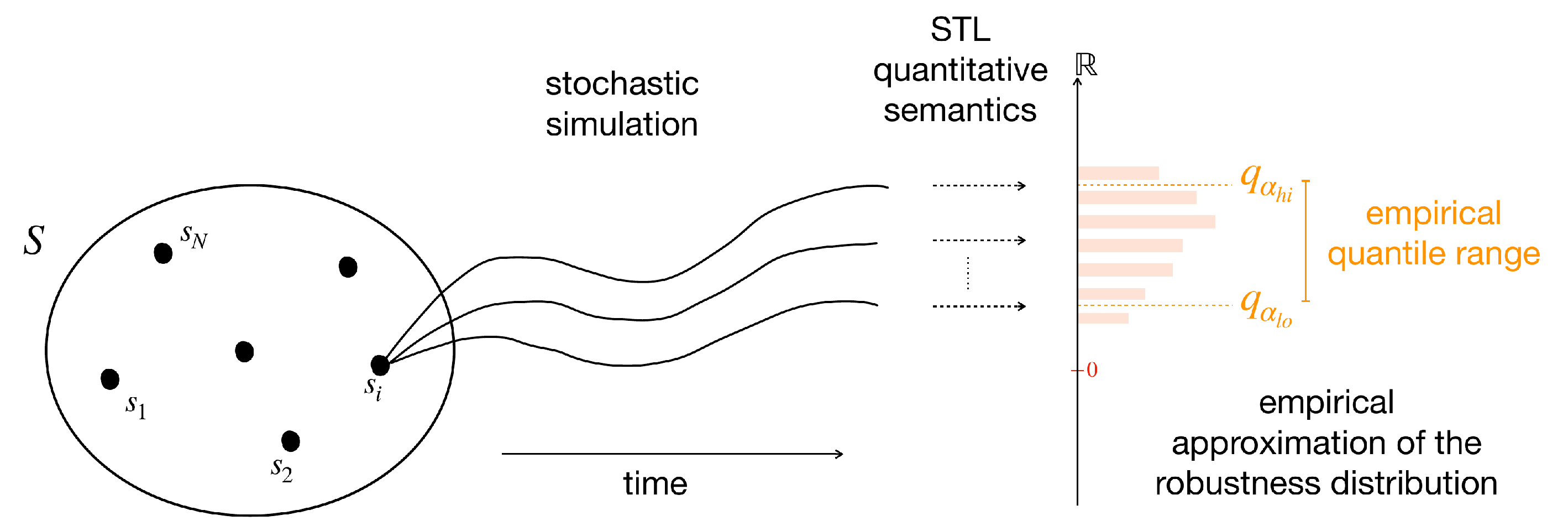}
              \caption{Diagram illustrating the generation of the dataset.\vspace{-0.5cm}}
              \label{fig:dataset_diagram}
          \end{figure}


\paragraph{Dataset generation.} In this step, we collect data for training the QR function and constructing the calibration set. To do so, we perform Monte-Carlo simulations of the process in order to obtain an empirical approximation of $\mathcal{R}_\phi(s)$. 

In particular, we randomly sample $N$ states $s_1,\ldots,s_N\sim \mathbf{S}(\cdot,\cdot)$. Then, for each state $s_i$, we simulate $M$ trajectories of length $H$, $\Vec{s}^1_i,\ldots , \Vec{s}^M_i$ where $\Vec{s}^j_i$ is a realization of $\mathbb{P}(\Vec{s} \mid \Vec{s}(k) = s_i)$, and compute the robustness value $R_\phi(\Vec{s}^j_i,0)$ of each of these trajectories. 
We note that $\{R_\phi(\Vec{s}^j_i)\}_{j=1}^M$ is an empirical approximation of $\mathcal{R}_\phi(s_i)$. 

Moreover, for a choice of $\alpha \in (0,1)$, we derive empirical quantiles $q_{\alpha/2}^i$ and $q_{1-\alpha/2}^i$ from samples $R_\phi(\Vec{s}^1_i),\ldots,R_\phi(\Vec{s}^M_i)$ and use these to label the state into one of \textit{safe}, \textit{unsafe}, or \textit{risky}: if the trajectories starting from $s_i$ satisfy $\phi$ with probability above $1-\alpha/2$ (i.e., $q_{\alpha/2}^i>0$), then we label $s_i$ as $\ell_i=+1$ (safe); if the probability of  satisfaction rather than violation here is below $\alpha/2$ (i.e., $q_{1-\alpha/2}^i<0$), then we label $s_i$ as $\ell_i=-1$ (unsafe); otherwise, we use label $\ell_i=0$ (risky).

The dataset is thus defined as 
\begin{equation}\label{eq:dataset}
    Z^\phi = \Big\{\Big(s_i,\big(R_\phi(\Vec{s}^1_i),\ldots,R_\phi(\Vec{s}^M_i)\big), \ell_i\Big), i=1,\ldots ,N \Big\}.
\end{equation}
Fig.~\ref{fig:dataset_diagram} shows an overview of the steps needed to generate the dataset.
  
The generation of the test set $Z^\phi_{test}$ is very similar to that of $Z^\phi$. The main difference is in that the number 
of trajectories that we simulate from each state $s$ is much larger than $M$. 
This allows us to obtain a highly accurate  
empirical approximation of the distribution of $\mathcal{R}_\phi(s)$, which we use as the ground-truth baseline in our experimental evaluation\footnote{In the limit of infinite sample size, the empirical approximation approaches the true distribution}.


\paragraph{QR training and residuals computation.} We divide the dataset $Z^\phi$ into a training set $Z^\phi_t$ and a calibration set $Z^\phi_c$. We then use $Z^\phi_t$ to train a QR that learns how to map states $s$ into three quantiles 

$$f(s;\hat{\theta})=\{\hat{q}_{\alpha_{lo}}(s), \hat{q}_{0.5}(s), \hat{q}_{\alpha_{hi}}(s)\},$$ 
where $\alpha_{lo} = \alpha/2$ and $\alpha_{hi} = 1-\alpha/2$. In order to better reconstruct the shape of the target distribution, we also predict the median quantile, $q_{0.5}(s)$. 

We then apply CQR, i.e. we compute the residuals of the QR over $Z^\phi_c$, as described in Equation~\ref{eq:CQR_cal_set} of the previous section, and find the critical residual value $\tau$.

\paragraph{Inference} For a test state $s_*$, this step involves predicting the relevant quantiles, $\hat{q}_{\alpha_{lo}}(s_*)$ and $\hat{q}_{\alpha_{hi}}(s_*)$, using the QR predictor and correct the resulting interval using the critical residual value $\tau$.
The calibrated interval returned at inference time becomes
$$CPI(s_*) := [\hat{q}_{\alpha_{lo}}(s_*)-\tau, \hat{q}_{\alpha_{hi}}(s_*)+\tau]\subseteq\mathbb{R}$$ and it is guaranteed to contain the true unknown robustness value with probability $(1-\alpha)$.

          \begin{figure}
              \centering
              \includegraphics[width=\columnwidth]{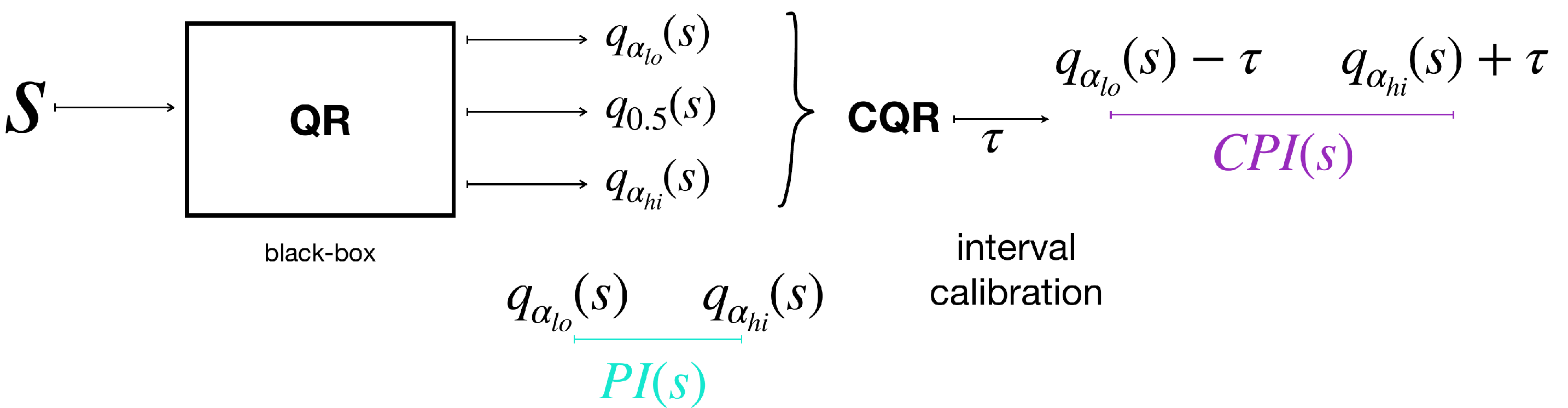}
              \caption{Overview of the conformal quantitative predictive monitoring technique.\vspace{-0.5cm}}
              \label{fig:cqpm_diagram}
          \end{figure}
          
\subsection{Composing monitors for composite properties}\label{subsec:modular}
The above-described QPM approach is end-to-end, in that it directly predicts the robustness of a fixed property $\phi$. On one hand, this is advantageous in that the monitor is tailored, and hence, highly accurate at predicting $R_{\phi}$ but on the other hand, it lacks flexibility. Thus, it becomes natural to ask oneself whether and how two predictive monitors trained for distinct properties $\phi_1$ and $\phi_2$ can be combined in a compositional manner to construct a monitor for a Boolean combination of the two properties, $\phi_1 \wedge \phi_2$ or $\phi_1 \vee \phi_2$, in such a way that the resulting monitor meets the desired probabilistic guarantees and doesn't require additional data collection or training. 

Here we propose two solutions to combine monitors for logical conjunction and disjunction. Handling the negation of a property, $\neg \phi$, is rather simple as we can take the interval with opposite signs. Therefore, we cover the basic Boolean logic. In what follows we will use a superscript $\phi_i$ with $i=1,2$ to distinguish predictors, datasets, critical values, and prediction regions of the two properties. 


\paragraph{\textbf{1. Union of intervals}.} The first, simple, solution considers the \emph{union} of the two property-specific prediction intervals. Mathematically, 
$$CPI^{\phi_{1,2}} := CPI^{\phi_{1}}\cup CPI^{\phi_{2}}.$$ 
We now prove that $CPI^{\phi_{1,2}}$ has a guaranteed coverage of $(1-\alpha$), the same of $CPI^{\phi_{1}}$ and $CPI^{\phi_{2}}$, for both $\phi_1\land\phi_2$ and $\phi_1\lor\phi_2$.

\begin{proposition}
    Given two properties $\phi_1$ and $\phi_2$, let $CPI^{\phi_i}$ be the monitor that maps each state $s\in S$ into a prediction interval for $\mathcal{R}_{\phi_i}(s)$ with coverage $1-\alpha$. Then, the interval $CPI^{\phi_{12}} := CPI^{\phi_{1}}\cup CPI^{\phi_{2}}$ is a valid prediction interval for both  $\mathcal{R}_{\phi_1\wedge\phi_2}$ and $\mathcal{R}_{\phi_1\vee\phi_2}$, i.e.,
    
    \begin{equation}
    \mathbb{P}\left(R_{\phi_1 \wedge \phi_2}(s) \in CPI^{\phi_{12}}(s)\right) \geq 1-\alpha
    \end{equation}
and
\begin{equation}
    \mathbb{P}\left(R_{\phi_1 \vee \phi_2}(s) \in CPI^{\phi_{12}}(s)\right) \geq 1-\alpha
    \end{equation}
    
for every state $s\in S$.

    \end{proposition}
The proof is provided in the Supplementary Material\footnote{Supplementary Material is available online in~\cite{cairoli2022conformal}.}.

\begin{figure}
              \centering
              \includegraphics[width=\columnwidth]{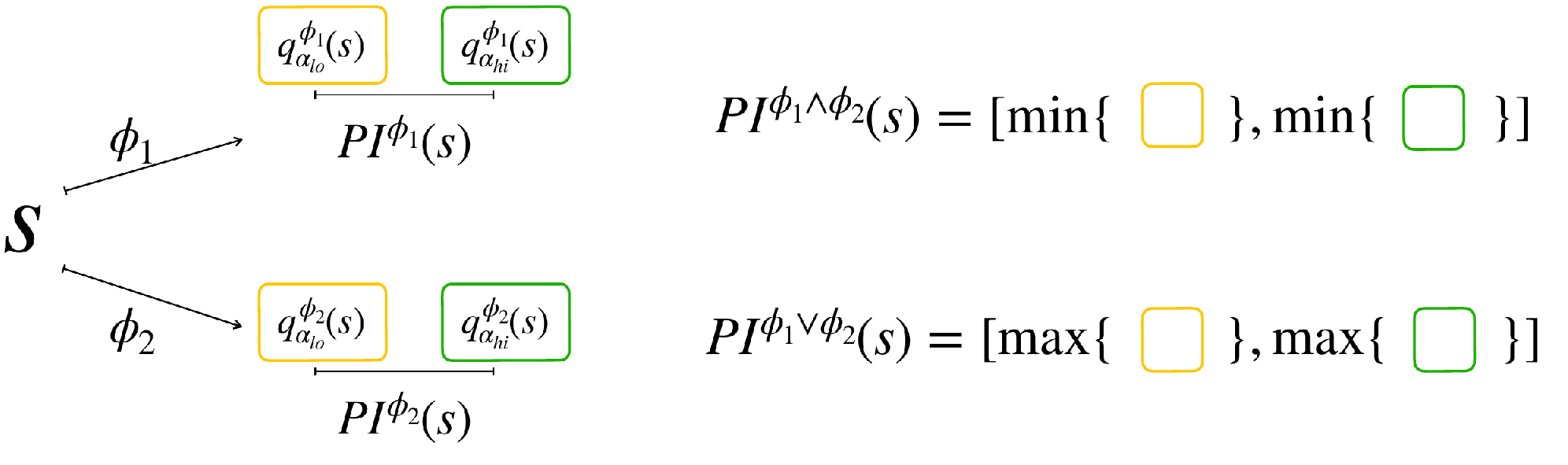}
              
              \vspace{-0.2cm}

              \caption{Overview of the combination of the conformal quantitative predictive monitors for two different properties $\phi_1$ and $\phi_2$.\vspace{-0.5cm}}
              \label{fig:comb_cqpm_diagram}
          \end{figure}

\paragraph{\textbf{2. Calibrated interval arithmetic}.} The second solution leverages the fact that CP (and CQR) can provide valid intervals on top of any predictor. Hence, we build a monitor for the composite property by first combining the predictors of the individual properties and then re-calibrating such obtained predictor. In particular, for  $\phi_{1}\land\phi_2$ we define for a state $s\in S$ the prediction interval as 
    \begin{equation*}
    PI^{\phi_{1}\land\phi_2}(s) :=
    \Big[\min\left(q^{\phi_1}_{\alpha_{lo}}(s),q^{\phi_2}_{\alpha_{lo}}(s)\right),\min\left(q^{\phi_1}_{\alpha_{hi}}(s),q^{\phi_2}_{\alpha_{hi}}(s)\right) \Big].\nonumber
    \end{equation*}
    See Fig.~\ref{fig:comb_cqpm_diagram} for a schematic visualization.
     We then compute the residuals of $PI^{\phi_{1}\land\phi_2}$ over the calibration set, obtaining the critical value $\tau^{\phi_1\land\phi_2}$ and the conformalized prediction interval
         \begin{align*}
    CPI^{\phi_1\land\phi_2}(s) :=
    \Big[\min&\left(q^{\phi_1}_{\alpha_{lo}}(s),q^{\phi_2}_{\alpha_{lo}}(s)\right)-\tau^{\phi_1\land\phi_2},\\
    &\min\left(q^{\phi_1}_{\alpha_{hi}}(s),q^{\phi_2}_{\alpha_{hi}}(s)\right)+\tau^{\phi_1\land\phi_2} \Big],\nonumber
    \end{align*}
    which has a guaranteed coverage of $1-\alpha$.
    
    The procedure is the same for property $\phi_{1}\lor\phi_2$, with the only difference that the prediction interval is given by 
    \begin{equation*}
    PI^{\phi_{1}\lor\phi_2}(s) :=
    \Big[\max\left(q^{\phi_1}_{\alpha_{lo}}(s),q^{\phi_2}_{\alpha_{lo}}(s)\right),\max\left(q^{\phi_1}_{\alpha_{hi}}(s),q^{\phi_2}_{\alpha_{hi}}(s)\right) \Big].\nonumber
    \end{equation*}
 
Solution (2) is expected to produce tighter intervals compared to solution (1)\footnote{In particular, the PI of solution (2) is strictly tighter than the CPI of solution (1).}. On the other hand, solution (2) requires the generation of the respective calibration set, $Z_c^{\phi_1\land\phi_2}$ or $Z_c^{\phi_1\lor\phi_2}$, which solution (1) does not need. 

\camera{
We remark that such compositionality is defined only for Boolean connectives and not for temporal operators.
}

\section{Experimental Results}\label{sec:experiments}
We experimentally evaluate the proposed QPM over a benchmark of four discrete-time stochastic processes with varying degrees of complexity. 

\subsection{Case Studies}\label{subsec:casestudies}
The first two case studies are selected from the stochastic hybrid benchmarks presented in the ARCH-COMP competitions~\cite{abate2021arch}. We then consider two stochastic hybrid systems with modular nature where the complexity can arbitrarily grow, making them suitable to test the scalability of the proposed solution. See the Supplementary Material (SM) for more details about the case studies presented below.

\paragraph{\textbf{Automated Anaesthesia Delivery}} (AAD)~\cite{abate2021arch} is a 3-dimensional discrete-time stochastic process with state $s(t) = (q(t), v(t))$ evolving according to linear dynamics with a controller and a Gaussian disturbance. The properties to be monitored are: $\phi_1 = G_{[0,t_H]}(v\in {\it Safe})$ and $\phi_2 = F_{[0,t_H]}(v_1 < v_2)$, where $\it Safe = [1, 6; 0, 10; 0, 10]$, $H = 60 min$, $\Delta t = 20 s$. The controller is defined as follows: $q = 7$ if $v_1 < 3.5$ and $q = 3.5$ if $v_1 \ge 3.5$.

\begin{figure}
    \centering
    \includegraphics[scale=0.54]{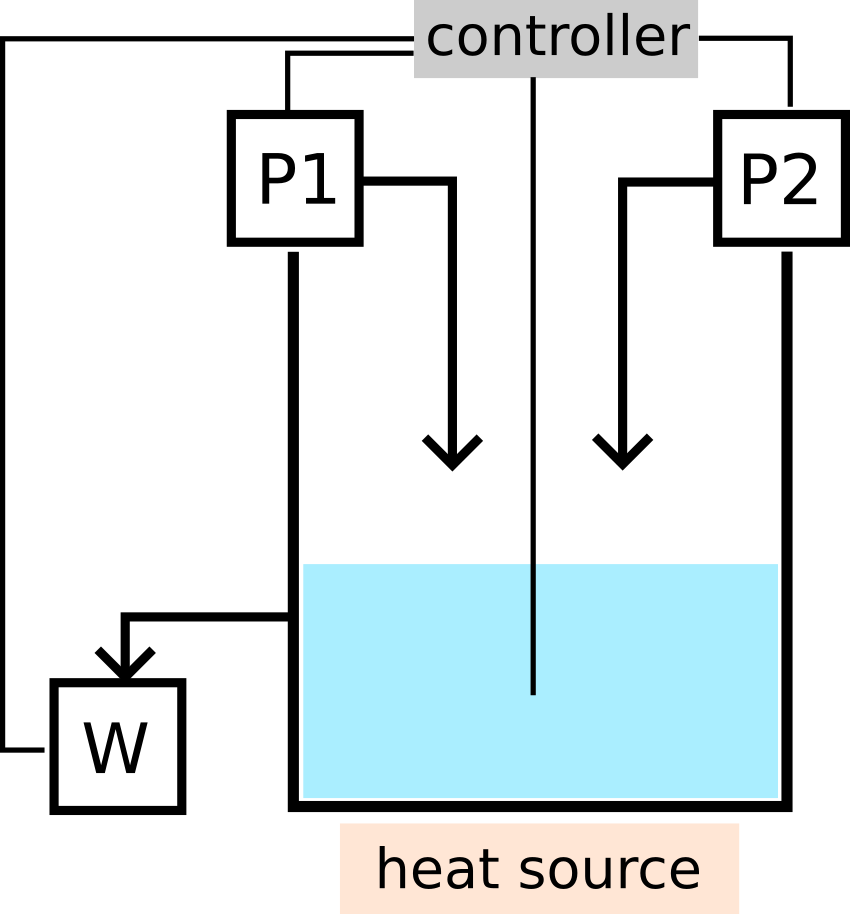}
    \caption{Scheme of the heated tank system whose components can undergo exponentially distributed failures.\vspace{-0.5cm}}
    \label{fig:ht_casestudy}
\end{figure}

\paragraph{\textbf{Heated Tank}} (HT) is a stochastic hybrid system composed of a tank containing a liquid whose level is influenced by two inflow pumps $P_1$ and $P_2$ and an outflow valve $W$, all managed by a controller $C$ (see Fig.~\ref{fig:ht_casestudy}). The liquid should absorb and transport the heat from a heat source. The discrete state space is five-dimensional. The continuous state space is two-dimensional and composed of a state $v(t) = (v_{height}(t), v_{temp}(t))$ representing the liquid height and the liquid temperature.
The general model adopted for this heated tank system is that of piece-wise deterministic Markov
processes, where the continuous part evolves according to switching differential equations. The switching is managed by the controller and by two exponentially distributed failure events. The property to be monitored is: the liquid level is always in the safe range $\it Safe = [H_{dryout},H_{overflow}]$ and the temperature is never too high and it can be expressed by the STL property
$ \phi = G_{[t_0,t_H]}\big((v_{height}\in {\it Safe})\land( v_{temp}\le T_{overheat})\big)$.

\begin{figure}[h]
\centering
\includegraphics[scale=0.36]{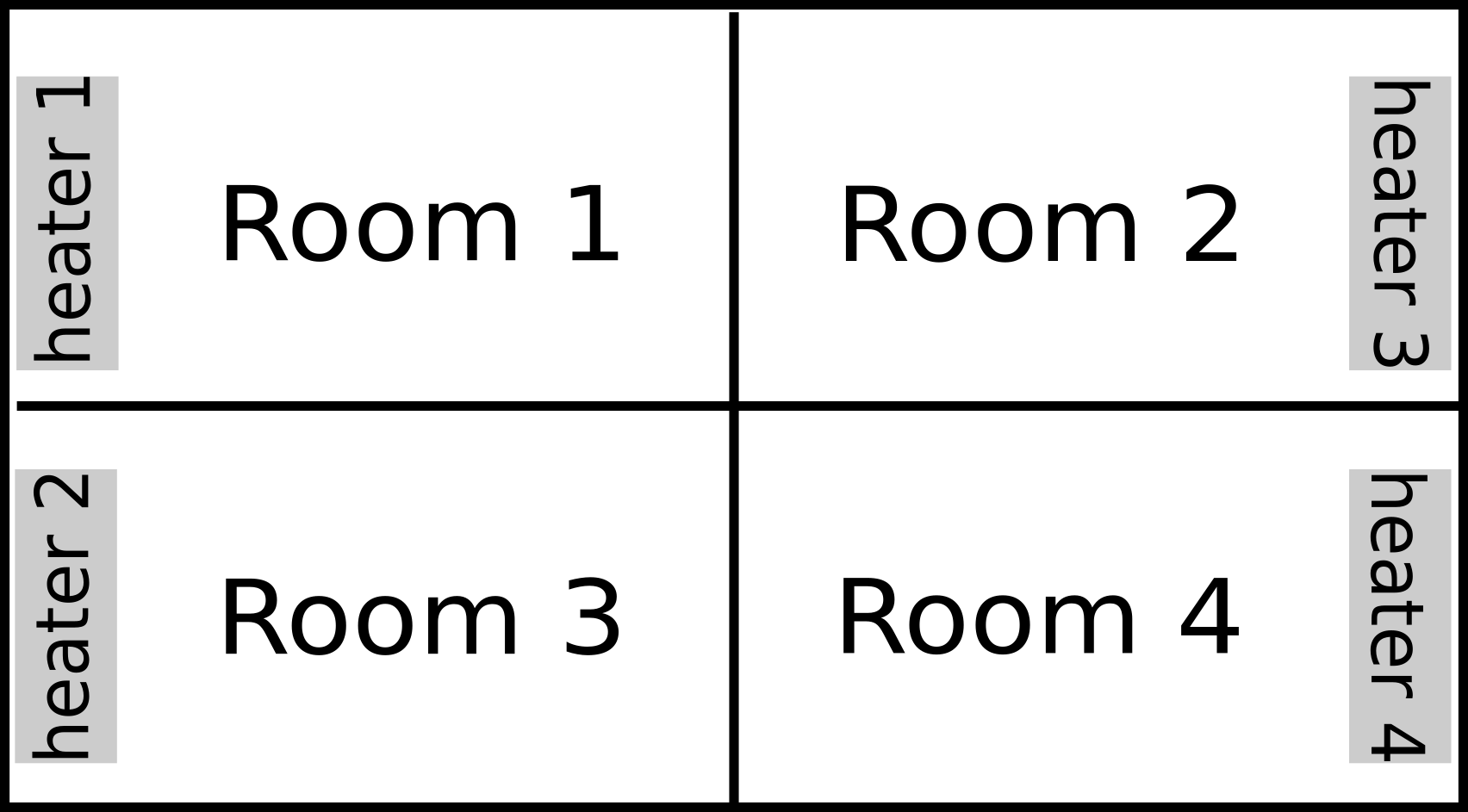}
\hspace{0.5cm}
\includegraphics[scale=0.72]{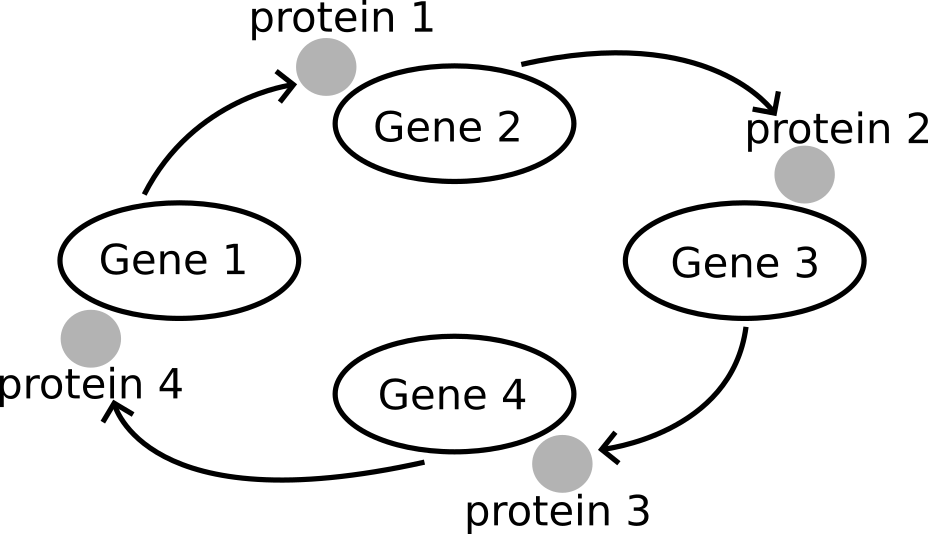}
\caption{Schematic visualization of the layout of the two modular case studies: multi-room heating (left) and gene regulatory network (right).\vspace{-0.5cm}}
        \label{fig:modular_casestudies}
    \end{figure}

\paragraph{\textbf{Multi-Room Heating}} (MRH)~\cite{abate2010approximate,fehnker2004benchmarks,malhame1985electric}: discrete-time stochastic hybrid system modeling the temperature evolution in a building with $h$ rooms. In each room is located a heater that switches between \emph{on} and \emph{off} depending on the current temperature in the room. The state of the system is hybrid $S= Q\times V$: the discrete state represents the status of the $h$ heaters, $Q=\{on,off\}^h$, and the continuous state represents the temperatures in the $h$ rooms, $V=\mathbb{R}^h$.  The average temperature evolves according to a stochastic difference equation that depends on the rooms' layout (see Fig.~\ref{fig:modular_casestudies} (left)). The room-specific requirement to be monitored is: the temperature $v_i$ of room $i$ always stays in a desirable range $[V_i^{lb},V_i^{ub}]$ and it can be expressed by the STL property $\phi_i = G_{[0,t_H]}(V_i^{lb}\le v_i\le V_i^{ub})$.

\paragraph{\textbf{Gene Regulatory Network}} (GRN)~\cite{bartocci2015system,kurasov2018stochastic}: discrete-time stochastic hybrid system modeling a genetic regulatory network composed of $h$ genes that produce respectively $h$ proteins that repress each other in a cyclic fashion (see Fig.~\ref{fig:modular_casestudies} (right)). Each gene could either be \emph{on}, actively producing its protein, or \emph{off}, expression is deactivated when a repressing protein binds to the gene. The state is thus hybrid $S= Q\times V$: the state of the genes is discrete, $Q=\{on,off\}^h$, whereas the proteins count is continuous, $V= \mathbb{N}^h$. 
Protein counts change deterministically according to a linear differential equation, whereas transitions between modes follow a Markov jump process. The gene-specific requirement to monitor is: the count $v_i$ of protein $i$ stabilize under a certain threshold $V_i^{ub}$ and it can be expressed by the STL property $\phi_i = G_{[t_{H/2},t_H]}(v_i\le V_i^{ub})$.
    
\subsection{Evaluation metrics}

We want our method to be capable of working at runtime in safety-critical applications,
which translates into the need for high reliability and high computational efficiency in producing the predictive intervals and in calibrating them. We emphasize that the time required to train the quantile regressor and to compute the calibration score $\tau$
does not affect its runtime efficiency, as it is performed in advance (offline) only once. Once trained, the
time needed to have a prediction interval for the current state is simply the
time needed to evaluate a neural network with a rather simple architecture. This time is almost negligible (in the order of
microseconds on GPU). 
In addition, we do not want an over-conservative predictor as an unnecessarily large interval would reduce the effectiveness of our QPM. 
Keeping that in mind, we introduce some relevant metrics to evaluate the performances of our QPM.

\paragraph{Accuracy}
We compare the sign of the predicted interval with the sign $\ell$ of the empirical quantile range over the test set. 
In simpler words, the latter can be interpreted as a classification accuracy over the sign of the prediction intervals and the dataset labels $\ell$. If both the real and the estimated interval are risky, i.e. if they both straddle zero, 
then the prediction is correct.
If the signs of the two intervals are opposite the prediction is \emph{wrong}. 
On the other hand, if the state is either safe or unsafe and it is predicted as risky, 
we label the prediction as \emph{uncertain}. 
Moreover, we analyze the percentage of \emph{false positive} (FP) errors, the most dangerous ones as they compromise the safety of the system. 
A false positive occurs when the state is either unsafe or risky but it is predicted to 
be safe.

\paragraph{Coverage and efficiency}
We experimentally check that the guaranteed validity of CQR is empirically met in the test evaluation. The efficiency represents the average width of the prediction intervals over the test set. In general, the larger the prediction interval the more conservative the CQR predictions. If the prediction intervals over the robustness values are always very large we have little information about the satisfaction of the property $\phi$. However, the predictive efficiency must be compared with the width of the empirical quantile range (EQR), i.e. the interval that contains $(1-\alpha)$ of the simulated robustness values. We can thus measure the conservativeness as the difference in width between the predicted efficiency and the EQR width. We also compare the coverage and the efficiency of QR against those of CQR.

Overall, we aim at reaching high accuracy in the reconstruction of the robustness distribution, which translates into a high percentage of correct predictions, i.e. intervals with agreeing signs. However, it is even more important to reduce the number of wrong predictions, especially false positives as they can compromise the safety of the system. 
We rather have more uncertain predictions but avoid erroneous ones. 
We check if the statistical guarantees about the coverage are met empirically by measuring the robustness of how many trajectories fall inside the predicted interval. 
Let us remark that the coverage and the percentage of correct predictions denote two different quantities \camera{and that the efficiency is not a percentage but the average width of the prediction interval measuring how conservative the predictor is compared to the empirical efficiency (EQR width)}. 

\subsection{Experiments}

 The workflow can be divided into steps: (1) define the model of the stochastic process, (2)  generate  the synthetic datasets $Z^\phi$ (simulate and compute STL robustness), (3) train the QR, (4) compute the calibration score $\tau$ (obtaining CQR prediction intervals), (5) evaluate  both the QR and the CQR on a test set $Z^\phi_{test}$.
 
\paragraph{Experimental settings} The entire pipeline is implemented in Python. The quantitative semantics of pcheck library\footnote{\url{https://github.com/simonesilvetti/pcheck}} is used to quantify the satisfaction of a certain formula for a specific trajectory. The
neural networks of the quantile regressors are trained with PyTorch~\cite{paszke2019pytorch}.
The experiments were conducted on a shared virtual machine with a 32-Core Processor, 64GB of RAM and an NVidia A100 GPU with 20GB, and 8 VCPU. 
Our implementation \camera{is available at \url{https://github.com/ailab-units/CQR_Quantitative_NPM}.}

\paragraph{Datasets} For every model, we generate a dataset $Z^\phi$ with size that increases with the dimensionality of the model. In general, the
chosen datasets are not too large. In particular, being $n$ the model's dimension, we sample $N_{train} =n\cdot 1000$ states for the training set, $N_{cal} = n\cdot 500$ states for the calibration set and $N_{test} = n\cdot 100$ states for the test set. From each state, we simulate $M=50$ trajectories for the training and for the calibration set and $M_{test} = 500$ trajectories for the test set. 
Data are scaled to the interval $[-1,1]$ to avoid sensitivity to different scales.

\paragraph{Training details} In all the experiments we use the same neural network architecture. The network is composed of $3$ hidden layers with $20$ neurons each. We use the Adam optimizer with a learning rate of $0.0005$ and a dropout rate of $0.1$ applied on each layer. We use a LeakyReLU activation function with slope $0.01$ and train the network for $500$ epochs over batches of size 512. 

\paragraph{Offline costs} The offline overhead consists of the time needed to generate the dataset and the time needed to train the neural network. The cost of the latter depends on the chosen architecture and on the dataset size. In our experiments, it ranges from $30$ minutes to $6$ hours for larger models. On the other hand, the time needed to generate the dataset depends on the time needed to simulate the trajectories and label them. These quantities are influenced by many factors: the chosen horizon $H$, the complexity of the STL property, and the complexity of the stochastic dynamics. For instance, the time needed to generate a dataset of $2000\times 50 = 100K=$ observations for the $GRN2$ model is double ($1$ hour) compared to the $MRH2$ model ($30$ minutes), even if they have the same dimensionality. The dataset generation time explodes easily around $20$ hour for $GRN6$ and $MRH8$. 

\paragraph{Performance evaluation} We choose in all the experiments $\alpha = 0.1$, so that $\alpha_{lo}=0.05$ and $\alpha_{hi} = 0.95$. The results are presented with the following structure. 
For each case study and for each property we train the QR and then recalibrate the predicted interval by means of CQR. 
For the modular case studies, i.e. MRH and GRN, we have a property regarding each individual component, i.e. for each room and for each gene respectively.
To analyze how the performance scales with the dimensionality of the underlying system, we test our QPM over different configurations of the modular system. In particular, we consider a multi-room heating (MRH) system with respectively 2, 4, and 8 rooms and a gene regulatory net (GRN) system with 2, 4, and 6 genes.

\begin{figure*}[!ht]
    \centering
        \includegraphics[width=.88\textwidth]{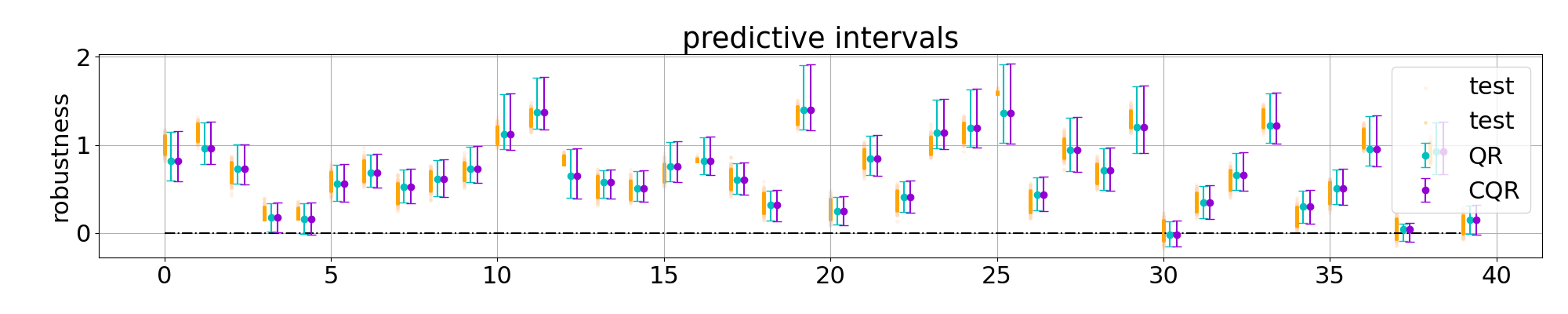}
        
        \vspace{-0.22cm}
        
        \includegraphics[width=.88\textwidth]{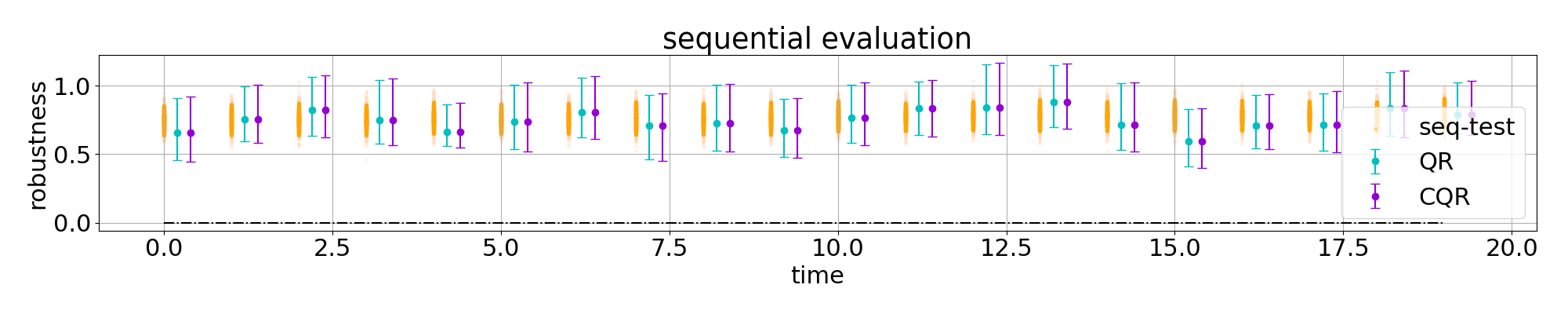}

\vspace{-0.4cm}

    \caption{Visualization of prediction intervals -- cyan for QR and blue for CQR -- over 40 randomly selected states (top) and for a runtime evaluation (bottom) for the automated anesthesia delivery system (\textbf{AAD-F}) w.r.t. property $\phi_2$ (eventually operator).\vspace{-0.25cm}
    }\label{fig:aad_plots}
\end{figure*}

\begin{table}

    \centering
\resizebox{\columnwidth}{!}{
    \begin{tabular}{l c c c c c c c }
        \toprule
         \textbf{AAD} & \textbf{correct} & \textbf{uncertain} & \textbf{wrong} & \textbf{FP} & \textbf{coverage} & \textbf{efficiency} &\textbf{EQR width} \\
         \midrule
QR & 83.00 & 14.50 & 2.50 & 2.50 & 87.86 & 0.1342 & 0.0497\\
CQR & 83.00 & 14.50 & 2.50 & 2.50 & 87.98 & 0.1349 &\\
\bottomrule
\toprule
         \textbf{AAD-F} & \textbf{correct} & \textbf{uncertain} & \textbf{wrong} & \textbf{FP} & \textbf{coverage} & \textbf{efficiency} &\camera{\textbf{EQR width}} \\
         \midrule
QR & 95.50 & 3.50 & 1.00 & 1.00 & 89.92 & 0.4333 & 0.1942\\
CQR & 95.50 & 3.50 & 1.00 & 1.00 & 91.28 & 0.4495 & \\
\bottomrule
\toprule
\textbf{HT}& \textbf{correct} & \textbf{uncertain} & \textbf{wrong} & \textbf{FP} & \textbf{coverage} & \textbf{efficiency} & \textbf{EQR width}   \\
         \hline
QR & 99.00 & 1.00  & 0.00 & 0.00 & 88.70 & 0.2778 & 0.2583 \\
CQR & 99.00 & 1.00  & 0.00 & 0.00 & 90.47 & 0.2812 & \\
\bottomrule

    \end{tabular}}
    \caption{Correct, uncertain, wrong and coverage represent percentage values ($\%$). \emph{EQR width} denotes the average width of the empirical quantile range over the test set.\vspace{-0.75cm}}
    \label{tab:res}
\end{table}

\begin{table}

    \centering
\resizebox{\columnwidth}{!}{
    \begin{tabular}{l c c c c c c c}
        \toprule
         \textbf{MRH2} & \textbf{correct} & \textbf{uncertain} & \textbf{wrong} & \textbf{FP} & \textbf{coverage} & \textbf{efficiency} & \textbf{EQR width}  \\
         \midrule
          \multicolumn{8}{c}{\textbf{Room 1}}\\
         \midrule
         
QR & 99.50  & 0.00 & 0.50& 0.00 & 89.25 & 0.3221 & 0.3167\\
CQR & 99.50  & 0.00 & 0.50& 0.00 & 90.19 & 0.3313 & \\
\midrule
         \multicolumn{8}{c}{\textbf{Room 2}}\\
         \midrule
QR & 86.50 & 11.00 & 2.50& 2.50 & 90.94 & 0.2568 & 0.1644\\
CQR & 86.50 & 11.00 & 2.50 & 2.50 & 91.58 & 0.2591 &\\
\midrule
\multicolumn{8}{c}{\textbf{Room 1 $\land$ Room 2}}\\
\midrule
QR & 99.50 & 0.00 & 0.50 &0.00 & 89.40 &0.3137 & 0.3026\\
CQR & 99.50& 0.00 & 0.50 & 0.00 & 90.33 &0.3225 &\\
MIN & 99.50 & 0.00 & 0.50 & 0.0 & 90.64 & 0.3230 &\\
UNION & - & - & - & - & 93.17 & 0.3169 & \\
\bottomrule

\toprule
         \textbf{MRH4} & \textbf{correct} & \textbf{uncertain} & \textbf{wrong} & \textbf{FP} & \textbf{coverage} & \textbf{efficiency} & \camera{\textbf{EQR width}}  \\
         \midrule
         \multicolumn{8}{c}{\textbf{Room 1}}\\
         \midrule
QR & 100.00 & 0.00 & 0.00 & 0.00 & 89.71 & 0.2945 & 0.2788\\
CQR & 100.00 & 0.00 & 0.00 & 0.00 & 90.03 & 0.2967 &\\
\midrule
         \multicolumn{8}{c}{\textbf{Room 2}}\\
         \midrule
QR & 91.00 & 6.75 & 2.25 & 2.25 & 90.10 & 0.2129 & 0.1540\\
CQR & 91.00 & 6.75 & 2.25 & 2.25 & 90.31 & 0.2151 &\\
\midrule
         \multicolumn{8}{c}{\textbf{Room 3}}\\
         \midrule
QR & 94.50 & 4.75 & 0.75& 0.75 & 91.21 & 0.2170 & 0.1707\\
CQR & 94.50 & 4.25 & 1.25& 1.00 & 90.40 & 0.2117 &\\
\midrule
         \multicolumn{8}{c}{\textbf{Room 4}} \\
         \midrule
QR & 100.00 & 0.00 & 0.00 & 0.00 & 90.70 & 0.3019  & 0.2766\\
CQR & 100.00 & 0.00 & 0.00 & 0.00 & 90.14 & 0.2970 &\\
\midrule

\multicolumn{8}{c}{\textbf{Room 1 $\land$ Room 2}}\\
\midrule
QR & 100.00 & 0.00 & 0.00 & 0.00 & 90.19 & 0.2820 & 0.2547\\
CQR & 100.00  &0.00 & 0.00&0.00 &90.70 &0.2861  &\\
MIN & 100.00 & 0.00 &0.00& 0.00 &90.13 & 0.2747  &\\
UNION & - & - & - & - & 93.50 & 0.2815 &\\
\midrule
\multicolumn{8}{c}{\textbf{Room 1 $\land$ Room 3}}\\
\midrule
QR & 100.00 & 0.00&0.00 &0.00 &89.57 &0.2818  & 0.2584\\
CQR & 100.00& 0.00 &0.00 &0.00 & 90.04& 0.2860  & \\
MIN & 100.00 & 0.00 & 0.00 & 0.00 & 90.17 & 0.2764 & \\
UNION & - & - & - & -& 93.96 & 0.2850 &\\
\midrule
\multicolumn{8}{c}{\textbf{Room 1 $\land$ Room 4}}\\
\midrule
QR & 100.00 & 0.00&0.00 &0.00 &  90.73 & 0.3016 & 0.2760\\
CQR & 100.00 & 0.00&0.00 &0.00 &  90.16& 0.2965 & \\

MIN & 100.00 & 0.00 &0.00 & 0.00  & 89.59 & 0.2938 &\\
UNION & - & - & - & - &91.79 & 0.3336 &\\

\bottomrule
\end{tabular}}
\caption{Multi room heating results with either two or four rooms. Correct, uncertain, wrong and coverage represent percentage values ($\%$). \emph{EQR width} denotes the average width of the empirical quantile range over the test set.\vspace{-0.85cm}}\label{tab:mrh24_res}
\end{table}


\begin{table}[ht]

    \centering
\resizebox{\columnwidth}{!}{
    \begin{tabular}{l c c c c c c c}
        \toprule
         \textbf{GRN2} &\textbf{correct} & \textbf{uncertain} & \textbf{wrong} & \textbf{FP} & \textbf{coverage} & \textbf{efficiency} & \textbf{EQR width}  \\
         \midrule
         \multicolumn{8}{c}{\textbf{Gene 1}}\\
         \midrule
QR & 90.00 & 6.50 & 3.50 & 3.50 & 88.95 & 0.8551  & 0.7998\\
CQR & 89.50 & 7.00 & 3.50 & 3.50 & 89.94 & 0.8744 &\\
\midrule
         \multicolumn{8}{c}{\textbf{Gene 2}}\\
         \midrule
QR & 86.00 & 11.50 & 2.50 & 1.00 & 89.55 & 0.5773  & 0.4229\\
CQR & 86.00 & 11.50 & 2.50 & 1.00 & 89.59 & 0.5780 &\\
\midrule
         \multicolumn{8}{c}{\textbf{Gene 1 $\land$ Gene 2}}\\
         \midrule
QR & 81.50&16.50& 2.00 & 0.00 & 88.18 & 0.6332 & 0.5266\\
CQR & 81.50&16.50& 2.00 & 0.00 & 88.25 & 0.6345 &\\
MIN & 81.50&16.00& 2.50 & 2.00 & 89.02 & 0.6432 &\\
UNION & - & - & - & - &  93.88 & 0.6308 &\\
\bottomrule

\end{tabular}}
    \caption{Gene regulatory network results. Correct, uncertain, wrong $\land$ coverage represent percentage values ($\%$). \emph{EQR width} denotes the average width of the empirical quantile range over the test set. \vspace{-1.02cm}}
    \label{tab:grn2_res}
\end{table}
\begin{table}[ht]

    \centering
\resizebox{\columnwidth}{!}{
    \begin{tabular}{l c c c c c c c}
\toprule
         \textbf{GRN4} & \textbf{correct} & \textbf{uncertain} & \textbf{wrong} & \textbf{FP} & \textbf{coverage} & \textbf{efficiency} & \textbf{EQR width}  \\
         \midrule
         \multicolumn{8}{c}{\textbf{Gene 1}}\\
         \midrule
QR & 93.25 & 4.25 & 2.50 & 2.50 & 89.69 & 0.5713 & 0.5062\\
CQR & 93.00 & 4.50 & 2.50 & 2.50  & 90.00 & 0.5756 &\\
\midrule
         \multicolumn{8}{c}{\textbf{Gene 2}}\\
         \midrule
QR & 90.75 & 6.50 & 2.75 & 2.75 &89.49 & 0.5232& 0.4359\\
CQR &  90.75 & 6.50 & 2.75 & 2.75 &89.09 & 0.5187&\\
\midrule
         \multicolumn{8}{c}{\textbf{Gene 3}}\\
         \midrule
QR & 100.00 & 0.00 & 0.00 & 0.00 & 89.04 & 0.3759 & 0.2620\\
CQR & 100.00 & 0.00 & 0.00 & 0.00 & 90.74 & 0.3942 & \\
\midrule
         \multicolumn{8}{c}{\textbf{Gene 4}}\\
         \midrule
QR & 91.75 & 6.25 & 2.00 & 1.25 & 90.49 & 0.3877 & 0.2820\\
CQR & 91.75 & 6.25 & 2.00 & 1.25 & 90.46 & 0.3873 &\\
\midrule

\multicolumn{8}{c}{\textbf{Gene 1 $\land$ Gene 2}}\\
\midrule
QR & 85.50 & 10.75  &3.75  &3.50 & 89.13 &0.5299 & 0.4723\\
CQR & 85.50 & 11.00 & 3.50 & 3.25 & 89.43 & 0.5346  &\\
MIN & 86.50 & 9.00 & 4.50 & 4.00 & 89.23 & 0.5196 &\\
UNION & - & - & - & - & 91.44 & 0.5884 & \\
\midrule

\multicolumn{8}{c}{\textbf{Gene 1 $\land$ Gene 3}}\\
\midrule
QR &92.75 & 4.25  &3.00 & 3.00 & 89.22 & 0.5482 & 0.4763\\
CQR &  92.75 & 4.25  &3.00 & 3.00 &  89.30 & 0.5439 &\\
MIN & 93.25 & 4.25 & 2.50 & 2.50 & 89.30 & 0.5334  &\\
UNION & - & - & - & - & 91.40 & 0.5384 & \\
\midrule

\multicolumn{8}{c}{\textbf{Gene 1 $\land$ Gene 4}}\\
\midrule
QR & 86.00 & 10.00 & 4.00 & 2.50 & 88.22 &0.4857  & 0.4133\\
CQR & 85.25 & 10.75 & 4.00 & 2.75 & 89.06 & 0.4943 &\\

MIN & 85.50 & 10.75 & 3.75 & 3.25 & 89.37 & 0.4708  &\\
UNION & - & - & - & - & 90.60 & 0.5289 & \\

\bottomrule

\end{tabular}}

\vspace{-0.5cm}

    \caption{Caption as in Table~\ref{tab:grn2_res}. \vspace{-1.1cm}}
    \label{tab:grn4_res}
\end{table}

\subsubsection{Results}

Fig.~\ref{fig:aad_plots} (additional figures in SM) 
shows a comparison of the predicted intervals, both PI (cyan) and CPI (purple), over 40 states randomly sampled from the test set and compare them with the empirical samples (orange) collected in the test set. Intervals above zero mean that the state is safe, intervals below mean that the state is unsafe, whereas intervals that straddle zero mean that the state is risky. Empirical samples that lie outside the EQR are denoted in a lighter orange.

Tables~\ref{tab:res}-\ref{tab:grn4_res} (additional tables in SM) 
provide a detailed summary of the performance and of the generalization capabilities of our QPM method. In particular, they compare the accuracy (rate of correct, uncertain, wrong and false positive predictions), the coverage and the efficiency of our solution over all the different case studies.

We observe that the percentage of wrong and FP predictions is extremely low in all the experimental configurations (always lower than $4\%$). The percentage of uncertain prediction interval is, in general, very low (lower than $10\%$). Some case studies make an exception and present a high percentage of uncertain predictions (around $30\%$ of the test predictions). This happens because most of the empirical quantile ranges in the test set have one of the two extremes very close to zero (see 
SM for an intuitive visualization of the problem).
Therefore, even if numerical results may seem to suggest that we have a poor reconstruction, in reality, we are successfully capturing the fact that most of the states are on the verge of violating the property.

From Tables~\ref{tab:res}-~\ref{tab:grn4_res} (and SM) we can also experimentally observe that CQR always meets the statistical guarantees: when QR is over-conservative CQR refines the predicted interval (negative $\tau$), whereas when QR is over-confident CQR enlarge the predicted interval (positive $\tau$). That is, the conformal recalibration adjusts the width of the interval w.r.t. desired confidence level, reducing the number of over-conservative predictions.
The efficiencies of the PI and CPI predicted intervals are compared to the width of the empirical quantile range computed over the test set (\emph{EQR width} column). CQR produces intervals that are on average $34\%$ more conservative than the EQR, with a minimum of $0.3\%$ and a maximum of $170\%$ (for the $AAD$). Recall that the efficiency is the width of the interval and so lower values are in general more desirable.

The modular experiments, $MRH2,MRH4,MRH8$, $GRN2,GRN4$ and $GRN6$, allow us to analyse the compositionality of QPM over the conjunction of properties (each property is room- or gene-specific). Due to space limitations, we focus only on the conjunction  ($\land$ operator). The same analysis can be done for the disjunction ($\lor$ operator). In Tables~\ref{tab:res}-~\ref{tab:grn4_res} (and SM), we compare the performances of four different approaches: QR and CQR trained over the training set $Z_t^{\phi_1\land\phi_2}$ and calibrated over $Z_c^{\phi_1\land\phi_2}$, the minimum w.r.t\. interval arithmetic (MIN in tables) of the PI intervals calibrated once again over $Z_c^{\phi_1\land\phi_2}$ (solution  $(2)$ presented in Sec.~\ref{subsec:modular}), and finally the union (UNION in tables) of the CPI intervals (solution $(1)$ presented in Sec.~\ref{subsec:modular}). Results confirm our intuition that the UNION approach produces over-conservative predictions, i.e. intervals with high coverage and high efficiency. The MIN approach, on the other hand, provides better coverage (less conservative) and better efficiency (comparable to that of CQR). In particular, the CQR approach produces intervals that are on average $12.50\%$ more conservative than the EQR, MIN's intervals are on average $10\%$ more conservative than the EQR, and UNION's intervals are on average $19\%$ more conservative, i.e. the intervals are wider.

\paragraph{Sequential data} All the results presented so far consider a dataset of states independently sampled from a distribution $\mathcal{S}$. By doing so, we aim at quantifying the generalization capabilities of our QPM. However, we are also interested in applying our QPM at runtime to systems that are evolving in time. States will thus have a temporal dependency, meaning that we lose the exchangeability requirement behind the theoretical validity of the CPI.
Fig.~\ref{fig:aad_plots} (bottom) -- additional figures in SM --
shows the performance of QPM applied to such sequential data. In particular, we randomly sampled a trajectory of length $20$ and applied the QPM to each state in the trajectory. We observe that QPM well captures the evolution of the robustness values but it tends to be either slightly over-confident or slightly over-conservative. This behaviour is typical of CP when the data-generating distribution at test time differs from the one used during the training and the calibration phase.

\subsection{Discussion}

In summary, we can see that the proposed QPM is effective in tackling the scalability issues surrounding the predictive problem for stochastic processes. The solution is inevitably approximate, as an exact solution is unfeasible even for extremely simple stochastic processes. In this sense, the statistical guarantees our method provides play a crucial role when dealing with safety-critical applications. We show that QPM performs reasonably well also at runtime with sequentially generated data. We plan to further investigate the theoretical guarantees of QPM when applied to sequential and non-exchangeable settings. 
We stress that the proposed methodology is not strictly related to the chosen measure of robustness. For instance, one could decide to focus on a notion of time robustness~\cite{lindemann2022temporal} and apply the same methodology described here for spatial robustness.

\section{Related Work}\label{sec:related}

Learning-based approaches to runtime PM have been recently proposed, including the so-called Neural Predictive Monitoring (NPM) method~\cite{phan2018neural,bortolussi2019neural,bortolussi2021neural,cairoli2021neural,cairoli2022neural}. In NPM, neural networks are used to infer the Boolean satisfaction of a property and conformal prediction (CP) are used to provide statistical guarantees. However, NPM is restricted to reachability properties. NPM has been extended to support some source of stochasticity in the system: in~\cite{cairoli2021neural} they allow partial observability and noisy observations, whereas in~\cite{cairoli2022neural} the system dynamics are stochastic but the monitor only evaluates the Boolean satisfaction of some quantile trajectories, providing a limited understanding of the safety level of the current state. 
Predictive monitoring under partial observability is also analysed in~\cite{chou2020predictive},  where the authors combine Bayesian state estimation with pre-computed reach sets to reduce the runtime overhead. While their reachability bounds are certified, no correctness guarantees can be established for the estimation step.

Other learning-based approaches for reachability prediction of stochastic systems include~\cite{bortolussi2016smoothed,bortolussi2022stochastic,djeridane2006neural,royo2018classification,yel2020assured,granig2020weakness}. Of these,~\cite{yel2020assured} develop techniques to detect potential prediction errors using neural network verification methods~\cite{ivanov2019verisig}. These verification methods, however, do not scale well on large models and support only specific classes of neural networks. 
In~\cite{bortolussi2022stochastic}, the authors introduce an efficient approximate Bayesian monitor for the satisfaction of STL properties for stochastic systems. However, this work focuses, once again, on the Boolean satisfaction.

Various learning-based PM approaches for temporal logic properties~\cite{qin2019predictive,qin2020clairvoyant,ma2021predictive,yoon2021predictive,yu2022model,rodionova2021time} have been recently proposed. 
In particular,~\cite{qin2019predictive} provide (like we do) guaranteed prediction intervals, but (unlike our method) they are limited to ARMA/ARIMA models. Ma et al.~\cite{ma2021predictive} use uncertainty quantification with Bayesian RNNs to provide confidence guarantees. However, these models are, by nature, not well-calibrated (i.e., the model uncertainty does not reflect the observed one~\cite{kuleshov2018accurate}), making the resulting guarantees not theoretically valid. 

We contribute to the state of the art by developing a quantitative predictive monitoring method that offers good scalability, provides statistical guarantees, and supports stochastic systems with rich STL-based requirements. Another important novelty is to monitor and predict the quantitative STL satisfaction of a requirement rather than the Boolean one.

\section{Conclusions}\label{sec:conclusions}

We presented quantitative predictive monitoring (QPM), a technique to reliably monitor the evolution of a stochastic system at runtime. In particular, given a requirement expressed as an STL formula,  QPM quantifies how robustly this requirement is satisfied by means of a range of STL robustness values. This interval undergoes a principled recalibration that guarantees a desired level of coverage, i.e. the interval covers the exact STL robustness values with a given confidence. The proposed technique avoids expensive Monte-Carlo simulations at runtime by leveraging conformalized quantile regression. The resulting method has very little overhead during runtime execution. 

Our experimental evaluation demonstrates that we overall reach a high accuracy and low rate of prediction errors, and the statistical guarantees are always met. The conformal approach also adjusts the width of the interval w.r.t. desired confidence level, reducing the number of over-conservative predictions. These results support the claim of having an efficient and reliable QPM.

In future work, we intend to explore the effects of our QPM method over temporal, rather than spatial, STL robustness. We will investigate a possible dynamics-aware approach to inference. The latter should aim at limiting the inference only to an estimate of the system manifold, i.e. the region of the state space that is likely to be visited by the evolving stochastic process.

\begin{acks}
This work has been partially supported by the PRIN project ``SEDUCE'' n. 2017TWRCNB and by the ``REXASI-PRO'' H-EU project, call HORIZON-CL4-2021-HUMAN-01-01, Grant agreement ID: 101070028. This study was carried out within the PNRR research activities of the consortium iNEST (Interconnected North-Est Innovation Ecosystem) funded by the European Union Next-GenerationEU (Piano Nazionale di Ripresa e Resilienza (PNRR) – Missione 4 Componente 2, Investimento 1.5 – D.D. 1058 23/06/2022, ECS\_00000043). This manuscript reflects only the Authors’ views and opinions, neither the European Union nor the European Commission can be considered responsible for them.
\end{acks}


\balance
\bibliographystyle{ACM-Reference-Format}
\bibliography{biblio}

\clearpage

\begin{appendices}

\section{Case Studies}\label{sec:casestudies}
The first two case studies are selected from the stochastic hybrid benchmarks presented in the ARCH-COMP competitions~\cite{abate2021arch}. We then consider two stochastic hybrid systems with modular nature where the complexity can arbitrarily grow, making them suitable to test the scalability of the proposed solution.

\paragraph{\textbf{Automated Anaesthesia Delivery.}} (AAD)~\cite{abate2021arch} is a three dimensional discrete-time stochastic process with state $s(t) = (q(t), v(t))$ evolving according to the following dynamics:
        $$v(t_{k+1}) = A\cdot v(t_k)+ b\cdot q(t_k) + w(t_k),$$
        where $A, b$ are patient-specific parameters and there is a Gaussian disturbance $w(t)\sim\mathcal{N}(0,10^{-3}\mathbb{I}_3)$.
    The properties to be monitored are about the safety of the system, $\phi_1 = G_{[0,t_H]}(v\in {\it Safe})$, $\phi_2 = F_{[0,t_H]}(v_1 < v_2)$ where $\it Safe = [1, 6; 0, 10; 0, 10]$, $H = 60 min$, $\Delta t = 20 s$. The controller is defined as follows: $q = 7$ if $v_1 < 3.5$ and $q = 3.5$ if $v_1 \ge 3.5$.

\paragraph{\textbf{Heated Tank.}} (HT) is a system composed of a tank $T$ containing a liquid whose level is influenced by two inflow pumps $P_1$ and $P_2$ and a outflow valve $W$, all managed by a controller $C$. The liquid should absorb and transport the heat from a heat source. The discrete state space is five-dimensional composed of states $q$:
$$
q(t) = (q_{P_1}(t), q_{P_2}(t), q_{W}(t), q_{C}(t), q_T(t)).
$$
The general model adopted for this heated tank system is that of piece-wise deterministic Markov
processes.
The continuous state space is two-dimensional composed of a state $v(t) = (v_{height}(t), v_{temp}(t))$ representing the liquid height and the liquid temperature. The continuous part evolves according to the following switching differential equations:
\begin{align*}
\dot{v}_H(t) &= (q_{P_1}(t)+q_{P_2}(t)-q_{W}(t))\cdot g\\  
\dot{v}_T(t) &= \frac{((\chi_{P_1}(t)+\chi_{P_2}(t))\cdot (T_{in}-v_{temp}(t))\cdot g + E_{in})}{v_{height}(t)},
\end{align*}
where $q_i$ indicates if unit $i\in\{P_1,P_2, W\}$ is working $1$ or not $0$, $g$ is the flow parameter, $T_{in} = 15\deg C$ is the inflow temperature and $E_{in}= 1\tfrac{\deg C m}{h}$ is the heat source parameter.
Each unit can be in one of the four discrete modes $\it Q_U=\{On,Off,StuckOn, StuckOff\}$. The switching from $\it On$ to $\it Off$ and vice-versa is managed by the controller. In addition to this switching, there are two exponentially distributed failure switching possibilities: from $\it On$ to $\it StuckOn$ and from $\it Off$ to $\it StuckOff$ with rates $\lambda_{P_1} = 1/219$, $\lambda_{P_2} = 1/175$ and $\lambda_{W} = 1/320$. The controller mode $q_C$ switches between the following configurations $\it Q_C = \{Normal, Increase, Decrease\}$. If $\it Normal$ the controller
does not try to influence the pumps and valve. If $\it Increase$ the controller aims to
increase the height of the liquid in the tank, by switching both pumps on, and by switching
the valve off. If $\it Decrease$ the controller aims to decrease the height of the liquid in
the tank, by switching both pumps off, and by switching the valve on. The switching by the
controller has no effect on unit $i$ if it is in failure mode $\it StuckOn$ or $\it StuckOff$. If $v_{height}(t)\le H_{low}$ then the controller switches from $\it Normal$ or $\it Decrease$ to $\it Increase$. If $v_{height}(t)\ge H_{high}$ the controller switches from $\it Normal$ or $\it Increase$ to $\it Decrease$. $H_{low} = 6 m$, $H_{high} = 8 m$. The property to monitor is 
$   \phi = G_{[t_0,t_H]}\big((H_{dryout}\le v_{height}\le H_{overflow})\land
    \land( v_{temp}\le T_{overheat})\big)$.

    \begin{center}
        \includegraphics[scale=0.8]{imgs/heated_tank.png}
    \end{center}
\paragraph{\textbf{Multi-Room Heating.}} (MRH)~\cite{abate2010approximate,fehnker2004benchmarks,malhame1985electric}: discete-time stochastic hybrid system modeling the temperature evolution in a building with $h$ rooms. In each room is located a heater that switches between \emph{on} and \emph{off} depending on the current temperature in the room. The state of the system is hybrid $S= Q\times V$: the discrete state represents the status of the $h$ heaters, $Q=\{on,off\}^h$, and the continuous state represents the temperatures in the $h$ rooms, $V=\mathbb{R}^h$.  The average temperature evolves according to the following
    stochastic difference equation:
    \begin{align*}
       v_i(t_{k+1}) &= v_i(t_k)+b_i(x_a-v_i(t_k))+\\
       +\sum_{i\ne j} a_{ij}(v_j(t_k)&-v_i(t_k))+c_i\mathbb{I}_{Q_i}(q(t_k))+w_i(t_k),
    \end{align*}
    where $x_a$ is the ambient temperature (constant for the entire building) and $\mathbb{I}_{Q_i}(\cdot)$ is the indicator function of set $Q_i:= \{(q_1,\ldots,q_h)\in Q: q_i = on\}$. The quantities $b_i$, $a_{ij}$ and $c_i$ are non-negative constants representing respectively the average heat transfer rate from room $i$ to the ambient ($b_i$) and to room $j\ne i$ ($a_{ij}$) and the heat rate supplied to room $i$ by the heater in
room $i$ ($c_i$). $a_{ij}$ also reflects the rooms layout, e.g. $a_{ij}=0$ if room $i$ and $j$ are not adjacent.
     The overall complexity  of the stochastic hybrid system is determined by the number of rooms. The room-specific requirement to monitor is: the temperature $v_i$ of room $i$ always stays in a desirable range $[V_i^{lb},V_i^{ub}]$ and it can be expressed by the STL property $\phi_i = G_{[0,t_H]}(V_i^{lb}\le v_i\le V_i^{ub})$.
    
    \begin{center}
        \includegraphics[scale=0.6]{imgs/multiroomheating4.png}
    \end{center}
    
\paragraph{\textbf{Gene Regulatory Network.}} (GRN)~\cite{bartocci2015system,kurasov2018stochastic}: discrete-time stochastic hybrid system modeling a genetic regulatory network composed of $h$ genes that produces respectively $h$ proteins that repress each other in a cyclic fashion. Each gene could either be \emph{on}, actively producing its protein, or \emph{off}, expression is deactivate when a repressing protein binds to the gene. The state is thus hybrid $S= Q\times V$: the state of the genes is discrete, $Q=\{on,off\}^h$, whereas the proteins count is continuous, $V= \mathbb{N}^h$. 
Protein counts change deterministically according to a linear differential equation, whereas transition between modes follow a Markov jump process. The change of protein $i$ has the form $a_i -b_iv_i$ if $q_i = off$ and $c_i -d_iv_i$ if $q_i = on$,
where $a_i$ and $c_i$ are the production rates of protein $i$ if the gene is inactive or active, respectively. The respective degradation rate constants are given by $b_i$ and $d_i$. The binding rate between a protein $i-1$ and a gene $i$ is expressed as $k_b\cdot v_{i-1}q_i$, whereas the unbinding rate is constant $k_u$. The shooting time for this reactions is exponentially distributed.
The complexity of the stochastic hybrid system is determined by the number of genes. The gene-specific requirement to monitor is: the count $v_i$ of protein $i$ stabilize under a certain threshold $V_i^{ub}$ and it can be expressed by the STL property $\phi_i = G_{[t_{H/2},t_H]}(v_i\le V_i^{ub})$.
    
    \begin{center}
        \includegraphics[scale=1.2]{imgs/generegulatory4.png}
    \end{center}

\section{Proof of Proposition 1}
\begin{proposition}
    Given two properties $\phi_1$ and $\phi_2$, let $CPI^{\phi_i}$ be the monitor that maps each state $s\in S$ into a prediction interval for $\mathcal{R}_{\phi_i}(s)$ with coverage $1-\alpha$. Then, the interval $CPI^{\phi_{12}} := CPI^{\phi_{1}}\cup CPI^{\phi_{2}}$ is a valid prediction interval for both  $\mathcal{R}_{\phi_1\wedge\phi_2}$ and $\mathcal{R}_{\phi_1\vee\phi_2}$, i.e.,
    
    \begin{equation}
    \mathbb{P}\left(R_{\phi_1 \wedge \phi_2}(s) \in CPI^{\phi_{12}}(s)\right) \geq 1-\alpha
    \end{equation}
and
\begin{equation}
    \mathbb{P}\left(R_{\phi_1 \vee \phi_2}(s) \in CPI^{\phi_{12}}(s)\right) \geq 1-\alpha
    \end{equation}
    
for every state $s\in S$.

    \end{proposition}
    \begin{proof}
    We know that, for $i \in\{1,2\}$ and for each state $s\in S$, $\mathbb{P}\big(R_{\phi_i}(s)\in CPI^{\phi_i}(s) \big)\ge 1-\alpha$. We also know that, by the STL quantitative semantics, $$\mathcal{R}_{\phi_1\land\phi_2}(s) = \min \Big(\mathcal{R}_{\phi_1}(s),\mathcal{R}_{\phi_2}(s)
    \Big).$$
The following inequalities hold:
\begin{align*}
   &\mathbb{P}\Big(\min \big(\mathcal{R}_{\phi_1}(s),\mathcal{R}_{\phi_2}(s)) 
    \big) \in CPI^{\phi_{12}}(s)\Big) =\\ &\mathbb{P}\big(\mathcal{R}_{\phi_1}(s)\in CPI^{\phi_{12}}(s) \big)+\mathbb{P}\big(\mathcal{R}_{\phi_2}(s)\in CPI^{\phi_{12}}(s) \big)\\
    &-\mathbb{P}\big(\mathcal{R}_{\phi_1}(s)\in CPI^{\phi_{12}}(s), \mathcal{R}_{\phi_2}(s)\in CPI^{\phi_{12}}(s) \big).
\end{align*}
The first two addends are higher than $1-\alpha$ since $CPI^{\phi_i}(s)\subseteq CPI^{\phi_{12}}(s)$ for every $i\in \{1,2\}$. The last addend is always lower than $\mathbb{P}\Big(\min \big(\mathcal{R}_{\phi_1}(s),\mathcal{R}_{\phi_1}(s)) 
    \big)\Big)$. Thus the inequality becomes:
    \begin{align*}
        2\mathbb{P}\Big(\min \big(\mathcal{R}_{\phi_1}(s),\mathcal{R}_{\phi_2}(s)) 
    \big)\in CPI^{\phi_{12}}(s)\Big)\ge 2(1-\alpha),
    \end{align*}
    which proves the proposition.
    The proof for the $\phi_1\lor\phi_2$ is very similar, the main difference is that
    $$
    \mathcal{R}_{\phi_\lor\phi_2}(s) = \max \Big(\mathcal{R}_{\phi_1}(s),\mathcal{R}_{\phi_1}(s)
    \Big).$$
 
    \end{proof}   

\clearpage

\section{Additional Tables of Results}

Table~\ref{tab:grn6_res} and~\ref{tab:mrh8_res} present the performances of our QPM over the high-dimensional modular case studies: $GRN6$ and $MRH8$ respectively. They compare the accuracy (rate of correct, uncertain, wrong and false positive predictions), the coverage and the efficiency of the proposed solution over all the different case studies.

\begin{table}[h]

    \centering
\resizebox{\columnwidth}{!}{
    \begin{tabular}{l c c c c c c c}
        \toprule
         \textbf{GRN6} &\textbf{correct} & \textbf{uncertain} & \textbf{wrong} & \textbf{FP} & \textbf{coverage} & \textbf{efficiency} & \textbf{EQR width}  \\
         \midrule
         \multicolumn{8}{c}{\textbf{Gene 1}}\\
         \midrule
QR & 92.00 & 6.83 & 1.17  & 1.17 & 88.92 & 0.4397 & 0.3772\\
CQR & 91.50 & 7.50 & 1.00 & 1.00 & 90.05 & 0.4523 & \\
\midrule
         \multicolumn{8}{c}{\textbf{Gene 2}}\\
         \midrule
QR & 90.67 & 8.00 & 1.33 & 1.16& 89.02& 0.4251& 0.3502\\
CQR & 90.67 & 8.17 & 1.17 &1.17 &89.57 &0.4303 & \\
\midrule
         \multicolumn{8}{c}{\textbf{Gene 3}}\\
         \midrule
QR & 92.17 &6.50 &1.33 &1.17 &89.44 &0.4022 & 0.3237\\
CQR & 91.83& 6.83&1.33 &1.17 &90.13 &0.4091 & \\
\midrule
         \multicolumn{8}{c}{\textbf{Gene 4}}\\
         \midrule
QR & 88.83 & 8.67& 2.50& 2.33& 89.11 &0.3832 & 0.3000\\
CQR & 89.00& 8.83 &2.17 &2.17 &89.83 &0.3904 & \\
\midrule
         \multicolumn{8}{c}{\textbf{Gene 5}}\\
         \midrule
QR & 100.00 & 0.00 &0.00 &0.00 & 88.33 & 0.3030 & 0.2032 \\
CQR & 100.00 & 0.00 &0.00 &0.00 & 89.83 & 0.3166&  \\
\midrule
         \multicolumn{8}{c}{\textbf{Gene 6}}\\
         \midrule
QR & 91.33 &7.00 &1.67 &0.83 & 88.90 &0.3656 & 0.2745\\
CQR & 91.33 & 7.00&1.67 &0.83 &89.19 &0.3688 & \\
\midrule
         \multicolumn{8}{c}{\textbf{Gene 1 $\land$ Gene 2}}\\
         \midrule
QR & 89.17 & 8.50& 2.33& 2.17& 89.41 & 0.4419&0.3834\\
CQR & 88.83& 8.33& 2.83& 2.67& 89.12& 0.4385& \\
MIN & 86.67 & 11.00 & 2.33& 2.33&89.43 &0.4293 & \\
UNION & - & -&- &- & 91.37& 0.4864& \\
\midrule
         \multicolumn{8}{c}{\textbf{Gene 1 $\land$ Gene 3}}\\
         \midrule
QR & 85.33 & 11.33&3.33 & 3.17& 87.80& 0.4125 & 0.3522\\
CQR & 84.33 & 12.83 & 2.83&2.67 & 89.02 &0.4242 & \\
MIN & 86.83 &10.67 &2.50 &2.17 &89.64 &0.4023 & \\
UNION & - & -&- &- &92.67 &0.4674 & \\
\midrule
         \multicolumn{8}{c}{\textbf{Gene 1 $\land$ Gene 4}}\\
         \midrule
QR & 88.50 & 9.83& 1.67 & 1.50 & 90.02& 0.4074& 0.3373\\
CQR & 88.33 &10.00 &1.67 &1.50 &90.31 &0.4104 & \\
MIN & 87.00 & 11.00 & 2.00&2.00 &89.91 &0.3937 & \\
UNION & - & -&- &- & 92.86& 0.4595& \\
    \midrule    \multicolumn{8}{c}{\textbf{Gene 1 $\land$ Gene 5}}\\
         \midrule
QR & 93.17  &5.00 &1.83 &1.50 &87.81 &0.4210 & 0.3552\\
CQR & 92.67& 5.67& 1.67& 1.50& 89.28 &0.4344 &\\
MIN & 92.00 &6.83 &1.17 &1.17 & 90.10 & 0.4165& \\
UNION & - & -&- &- & 91.80&0.4295& \\
\bottomrule
        \multicolumn{8}{c}{\textbf{Gene 1 $\land$ Gene 6}}\\
         \midrule
QR & 86.00 & 11.83& 2.17& 1.17 & 89.53& 0.4216&0.3466\\
CQR & 85.83 & 12.17 &2.00 &1.00 &90.02 &0.4277 &\\
MIN & 87.00 &10.17 &2.83 &1.50 &89.47 &0.4003 & \\
UNION & - & -&- &- & 91.47&0.4468 & \\
\end{tabular}}
    \caption{Gene regulatory network results. Correct, uncertain, wrong $\land$ coverage represent percentage values ($\%$). \emph{EQR width} denotes the average width of the empirical quantile range over the test set.}
    \label{tab:grn6_res}
\end{table}

\begin{table}[ht]
    \centering
\resizebox{\columnwidth}{0.7\height}{
    \begin{tabular}{l c c c c c c c}
        \toprule
         \textbf{MRH8} & \textbf{correct} & \textbf{uncertain} & \textbf{wrong} & \textbf{FP} & \textbf{coverage} & \textbf{efficiency} & \textbf{EQR width}  \\
\midrule
\multicolumn{8}{c}{\textbf{Room 1}}\\
\midrule
QR & 67.38 & 28.75 & 3.88 & 3.75 & 89.04 & 0.2306 & 0.2219\\
CQR & 67.63  &31.13 & 1.25  &1.13 &89.87 &0.2346  &\\
\midrule
\multicolumn{8}{c}{\textbf{Room 2}}\\
\midrule
QR & 95.13 & 4.38 & 0.50 & 0.50 & 90.37 &0.1728  & 0.1289\\
CQR & 95.25  &4.25 &0.50  &0.50 &90.31& 0.1723 &\\
\midrule
\multicolumn{8}{c}{\textbf{Room 3}}\\
\midrule
QR & 95.88 & 3.50 & 0.63 & 3.75 & 88.55 &0.1621  & 0.1227\\
CQR & 95.50  & 3.88 & 0.63  &0.38 &90.35 &0.1670  &\\
\midrule
\multicolumn{8}{c}{\textbf{Room 4}}\\
\midrule
QR & 100.00 & 0.00 & 0.00 & 0.00  & 89.93 & 0.2593 & 0.2581\\
CQR &  100.00 & 0.00 & 0.00 & 0.00 & 90.13& 0.2607 &\\
\midrule
\multicolumn{8}{c}{\textbf{Room 5}}\\
\midrule
QR &  100.00 & 0.00 & 0.00 & 0.00  & 89.63 & 0.2527 &0.2529 \\
CQR &   100.00 & 0.00 & 0.00 & 0.00 & 89.98 & 0.2537 &\\
\midrule
\multicolumn{8}{c}{\textbf{Room 6}}\\
\midrule
QR & 95.88 & 3.00 & 1.13 & 1.13 & 89.72 & 0.1815 & 0.1407\\
CQR & 95.88  & 3.00 & 1.13 &1.13 &89.58 &0.1811  &\\
\midrule
\multicolumn{8}{c}{\textbf{Room 7}}\\
\midrule
QR & 95.88 & 3.75 & 3.75 & 0.25 & 89.84 &0.1645  & 0.1122\\
CQR &  95.88 & 3.75 & 3.75 & 0.25&  89.90 & 0.1652 &\\
\midrule

\multicolumn{8}{c}{\textbf{Room 8}}\\
\midrule
QR &  100.00 & 0.00 & 0.00 & 0.00  & 90.11 & 0.2531 & 0.2446\\
CQR &   100.00 & 0.00 & 0.00 & 0.00 & 89.99&0.2511  &\\
\midrule

\multicolumn{8}{c}{\textbf{Room 1 $\land$ Room 2}}\\
\midrule
QR & 69.25 & 30.75 & 0.00 & 0.00 & 89.77 & 0.2085 & 0.1897\\
CQR &  69.25 & 30.75 & 0.00 & 0.00&89.77 & 0.2085 &\\
MIN & 68.38 & 27.75 &3.88 &3.75  &90.16 & 0.2060  & \\
UNION & - & - & - & - & 92.82 & 0.2271 & \\

\midrule
\multicolumn{8}{c}{\textbf{Room 1 $\land$ Room 3}}\\
\midrule
QR & 69.25 & 30.75 & 0.00 & 0.00 & 90.05 & 0.2110 & 0.1905\\
CQR &  69.25 & 30.75 & 0.00 & 0.00&89.77 & 0.2085 &\\
MIN & 69.125 & 28.25 & 2.63 & 2.38 & 89.61 & 0.2069  & \\
UNION & - & - & - & - & 92.20 & 0.2240 & \\

\midrule
\multicolumn{8}{c}{\textbf{Room 1 $\land$ Room 4}}\\
\midrule
QR &  100.00 & 0.00 & 0.00 & 0.00 &  89.91& 0.2590 & 0.2580\\
CQR & 100.00 & 0.00 & 0.00 & 0.00 & 90.11 & 0.2605 &\\
MIN & 100.00 & 0.00 & 0.00 & 0.00  &90.07 & 0.2606  & \\
UNION & - & - & - & - & 90.14 & 0.2795 & \\

\midrule
\multicolumn{8}{c}{\textbf{Room 1 $\land$ Room 5}}\\
\midrule
QR & 100.00 & 0.00 & 0.00 & 0.00  & 89.63 & 0.2527 & 0.2529\\
CQR &  100.00 & 0.00 & 0.00 & 0.00 & 89.77 & 0.2537 &\\
MIN &100.00  & 0.00 & 0.00& 0.00 & 89.85& 0.2546  & \\
UNION & - & - & - & - & 89.88 & 0.2789 & \\

\midrule
\multicolumn{8}{c}{\textbf{Room 1 $\land$ Room 6}}\\
\midrule
QR & 68.13 & 29.88 & 2.00 & 2.00 & 89.62 & 0.2129 & 0.1945\\
CQR & 69.00  &30.38& 0.63 &0.63 &89.97 & 0.2141 &\\
MIN &67.38  & 28.00 & 4.63& 4.50 & 90.34& 0.2093  & \\
UNION & - & - & - & - & 92.80 & 0.2341 & \\

\midrule
\multicolumn{8}{c}{\textbf{Room 1 $\land$ Room 7}}\\
\midrule
QR & 68.50 & 30.75 & 0.75 & 0.75 & 89.27 & 68.50  & 0.1943\\
CQR & 69.00  &31.00 &0.00  &0.00 &89.95 & 0.2124  &\\
MIN & 68.25 & 28.13 & 3.63 & 3.50 & 89.19 & 0.2099  & \\
UNION & - & - & - & - & 91.90 & 0.2215 & \\
\midrule
\multicolumn{8}{c}{\textbf{Room 1 $\land$ Room 8}}\\
\midrule
QR & 99.63 & 0.13 & 0.25  & 0.00  & 89.35 & 0.2283 & 0.2217\\
CQR & 99.63 & 0.13 & 0.25  & 0.00 & 89.68 & 0.2306 &\\
MIN & 99.25 &0.13  &0.63 & 0.00 &90.12 & 0.2293  & \\
UNION & - & - & - & - & 94.29 & 0.2428 & \\

\bottomrule
    \end{tabular}}
    \caption{Multi room heating results with 8 rooms. Correct, uncertain, wrong and coverage represent percentage values ($\%$). \emph{EQR width} denotes the average width of the empirical quantile range over the test set.}\label{tab:mrh8_res}
\end{table}

\clearpage

\section{Additional Visualisations}

\begin{figure*}[t]
    \centering

        \includegraphics[width=\textwidth]{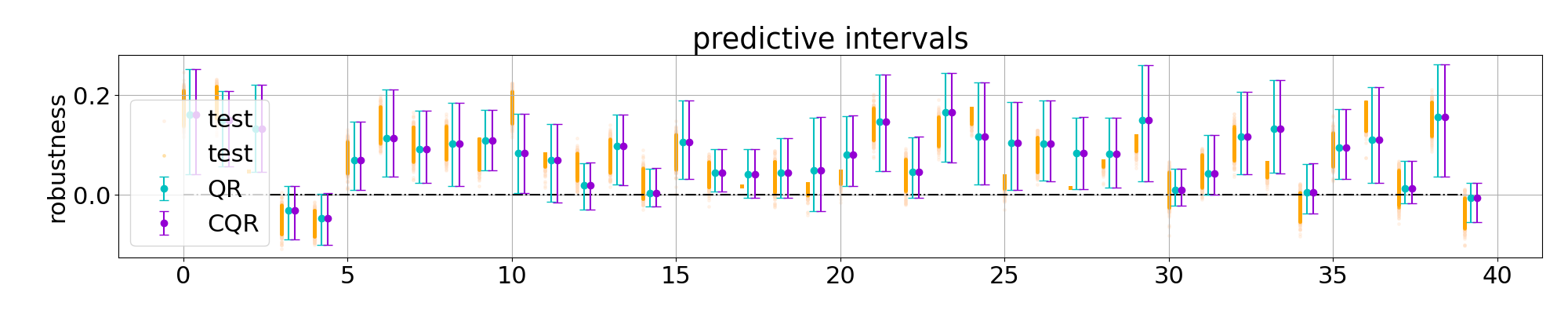}
    \caption{Visualization of prediction intervals -- cyan for QR and blue for CQR -- over 40 randomly selected states (top) and for a runtime evaluation (bottom) for the automated anaesthesia delivery system (\textbf{AAD}): property $\phi_1$ (with the globally operator). 
    }\label{fig:aad_g_plots}
\end{figure*}

\begin{figure*}
    \centering
        \includegraphics[width=\textwidth]{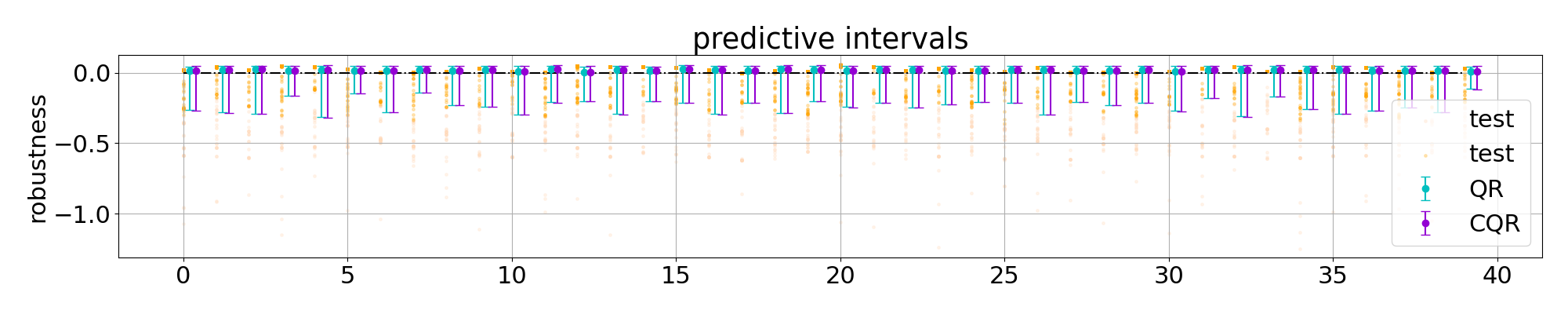}
        
    \caption{Visualization of prediction intervals -- cyan for QR and blue for CQR -- over 40 randomly selected states for the heated tank system (\textbf{HT}). The central dot denotes the predicted median.}
    \label{fig:ht_plots}
\end{figure*}

\begin{figure*}[ht]
    \centering
        \includegraphics[width=\textwidth]{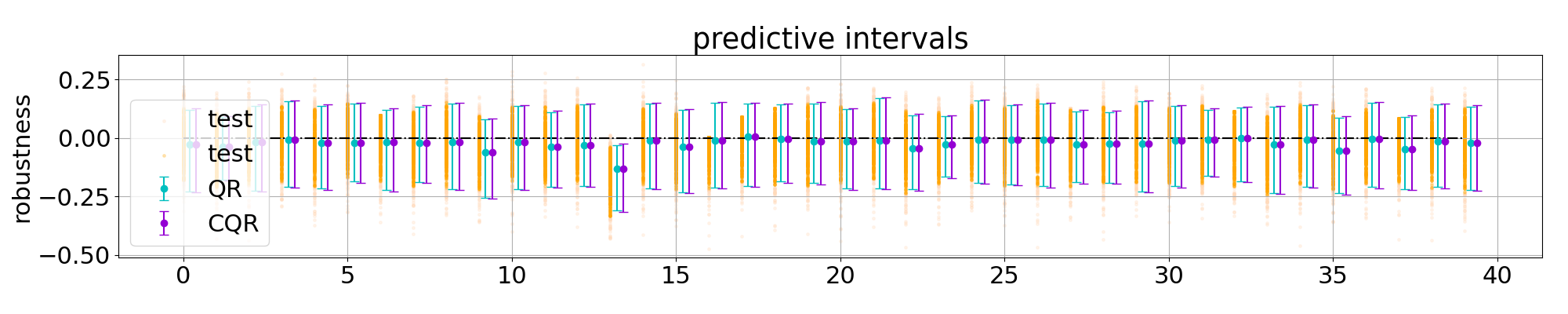}
        \includegraphics[width=\textwidth]{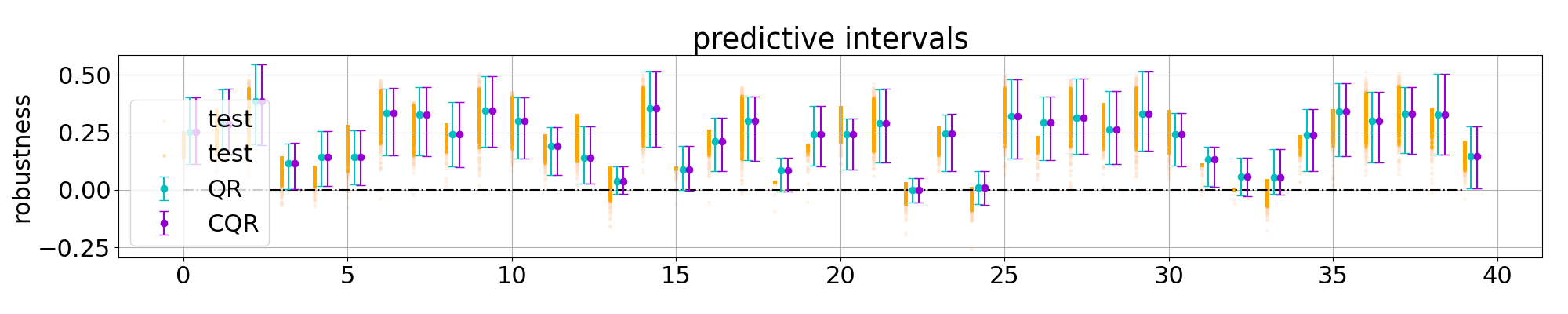}
    \caption{Visualization of prediction intervals -- cyan for QR and blue for CQR -- over 40 randomly selected states for a multi-room heating system with two rooms: property for room 1 (top) and property for room 2 (bottom) (\textbf{MRH2}). The central dot denotes the predicted median.}
    \label{fig:mrh2_plots}
\end{figure*}

\begin{figure*}[ht]
    \centering
        \includegraphics[width=\textwidth]{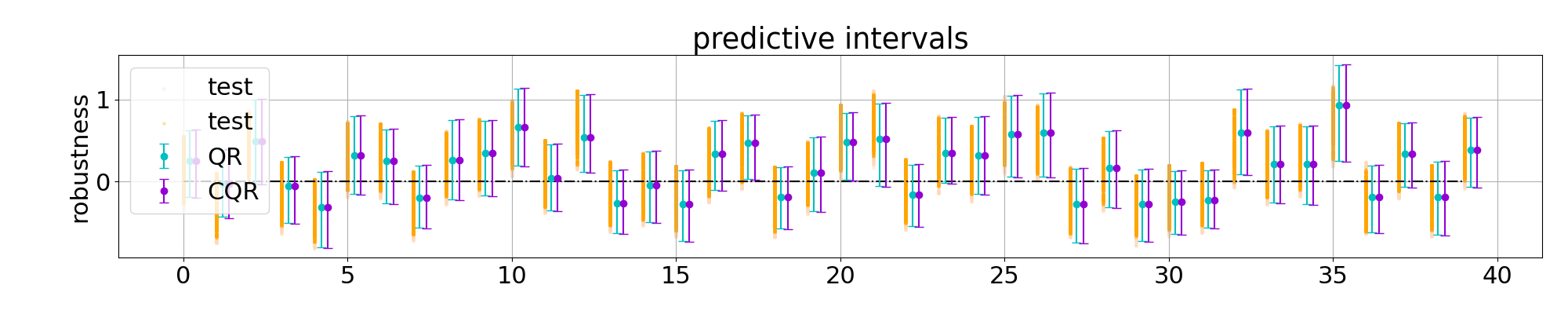}
        \includegraphics[width=\textwidth]{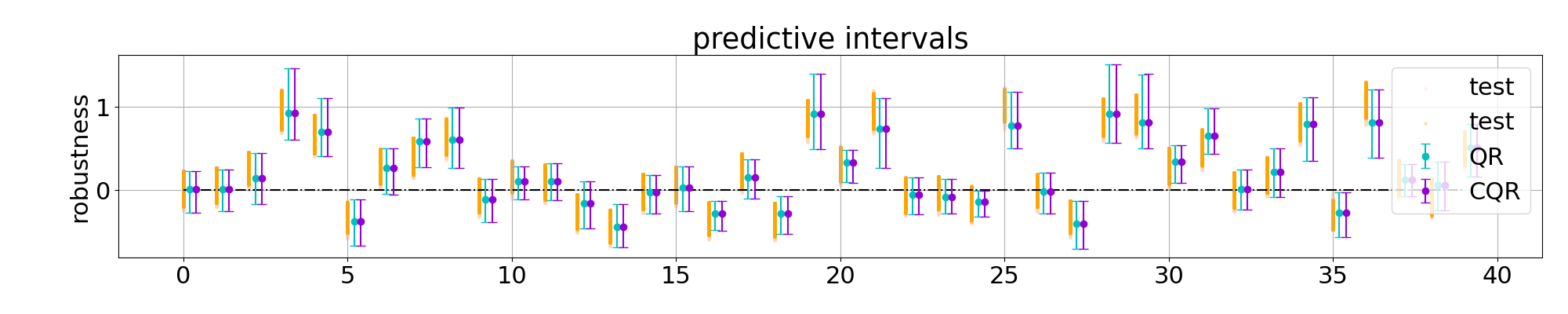}
    \caption{Visualization of prediction intervals -- cyan for QR and blue for CQR -- over 40 randomly selected states for a gene regulatory net system with two genes: property for gene 1 (top) and property for gene 2 (bottom) (\textbf{GRN2}). The central dot denotes the predicted median.}
    \label{fig:grn2_plots}
\end{figure*}

\begin{figure*}[ht]
    \centering
    \includegraphics[width=\textwidth]{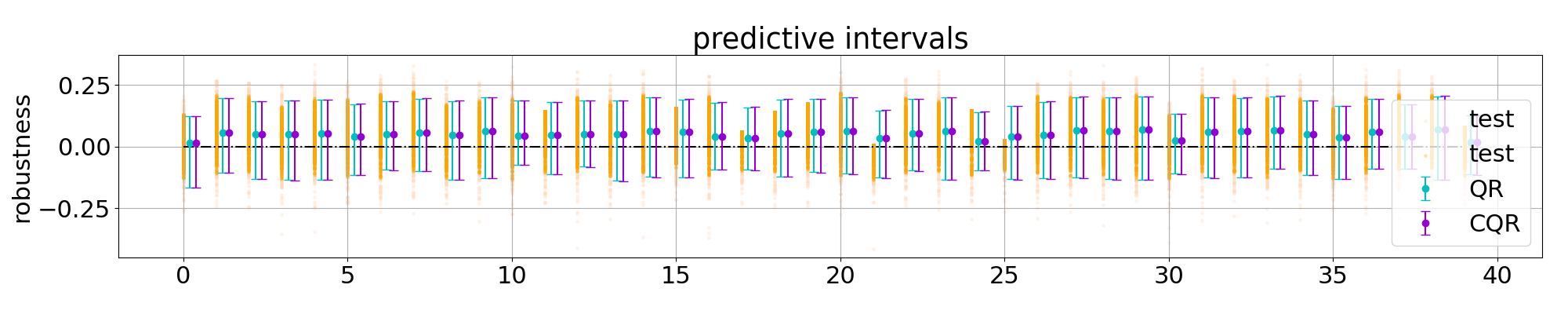}
    
    \caption{Visualization of prediction intervals -- cyan for QR and blue for CQR -- over 40 randomly selected states for a multi-room heating system with four rooms (\textbf{MRH4}): property for room 1. The central dot denotes the predicted median.}
    \label{fig:mrh4_plots}
\end{figure*}

\begin{figure*}[ht]
    \centering
    \includegraphics[width=\textwidth]{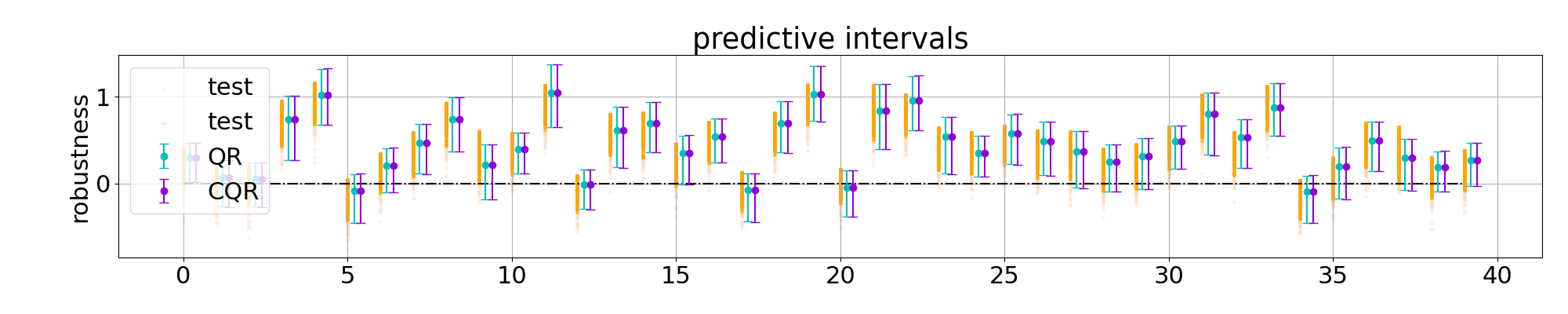}
    
    \caption{Visualization of prediction intervals -- cyan for QR and blue for CQR -- over 40 randomly selected states for a gene regulatory net system with four genes (\textbf{GRN4}): property for room 1. The central dot denotes the predicted median.}
    \label{fig:grn4_plots}
\end{figure*}

\begin{figure*}[ht]
    \centering
    \includegraphics[width=\textwidth]{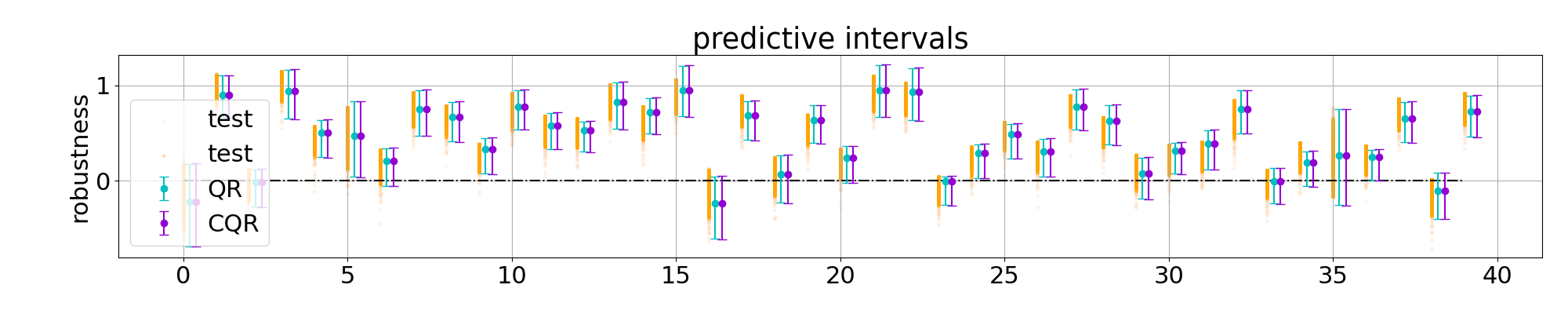}
    
    \caption{Visualization of prediction intervals -- cyan for QR and blue for CQR -- over 40 randomly selected states for a gene regulatory net system with six genes (\textbf{GRN6}): property for gene 1. The central dot denotes the predicted median.}
    \label{fig:grn6_plots}
\end{figure*}

\begin{figure*}[ht]
    \centering
    \includegraphics[width=\textwidth]{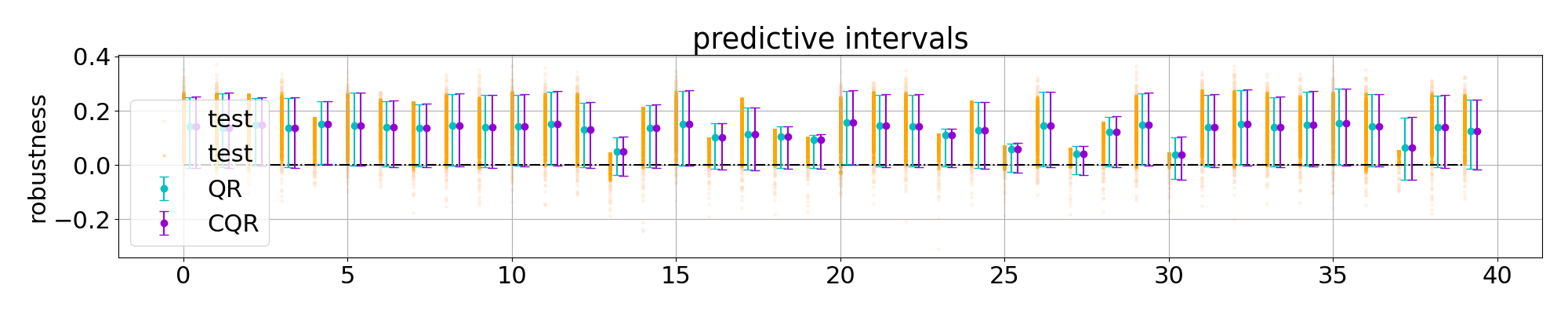}
    
    \caption{Visualization of prediction intervals -- cyan for QR and blue for CQR -- over 40 randomly selected states for a multi-room heating system with eight rooms (\textbf{MRH8}): property for room 1. The central dot denotes the predicted median.}
    \label{fig:mrh8_plots}
\end{figure*}

Fig.~\ref{fig:aad_g_plots}-\ref{fig:mrh8_plots} show a comparison of the predicted intervals, both PI (cyan) and CPI (purple), over 40 states randomly sampled from the test set and compare them with the empirical samples (orange) collected in the test set. Intervals above zero means that the state is safe, intervals below means that the state is unsafe, whereas intervals that straddle the zero means that the state is risky. Empirical samples that lie outside the EQR are denoted in a lighter orange.

\begin{figure*}
\centering

\includegraphics[width=\textwidth]{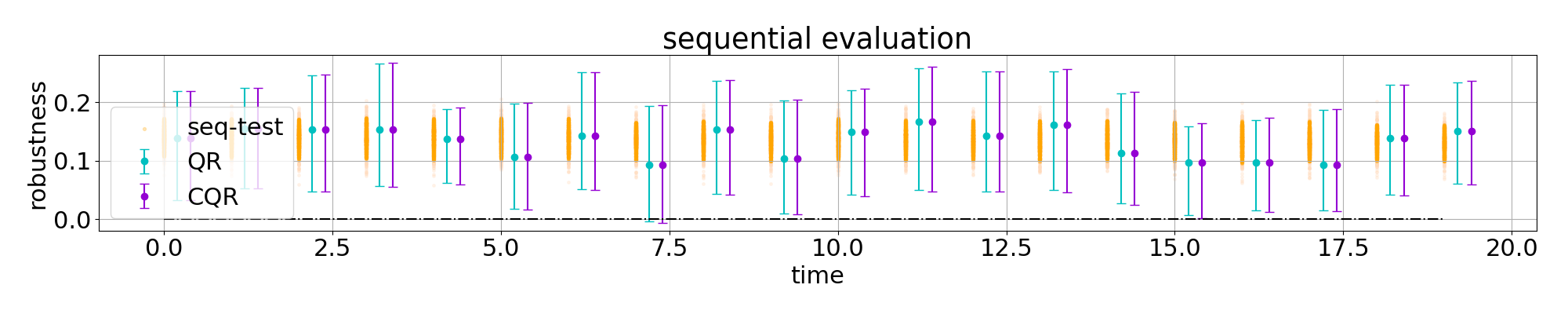}
\includegraphics[width=\textwidth]{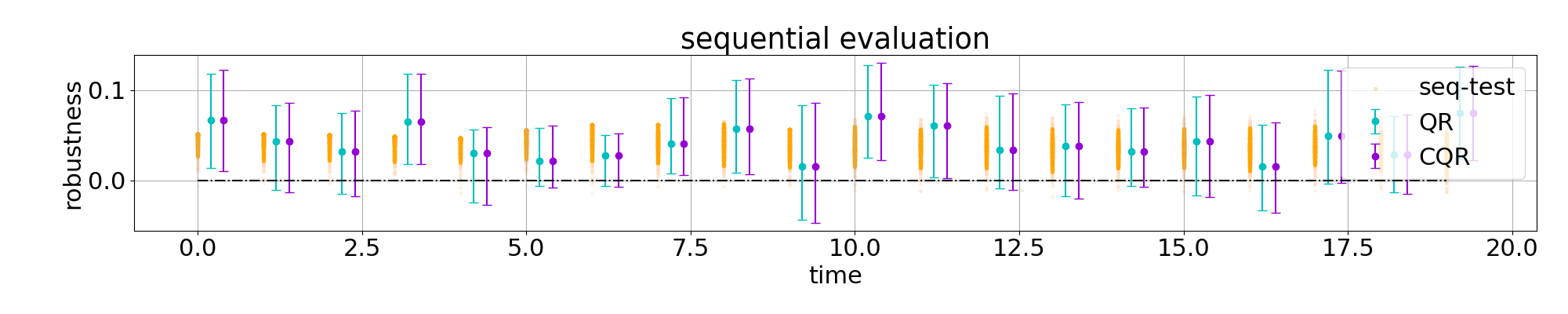}
\caption{Runtime evaluation of the AAD model w.r.t. property $\phi_1$ (\emph{eventually} operator) over two randomly selected initial states.
}\label{fig:aad_seq_plots}
\end{figure*}

\begin{figure*}
\centering

\includegraphics[width=\textwidth]{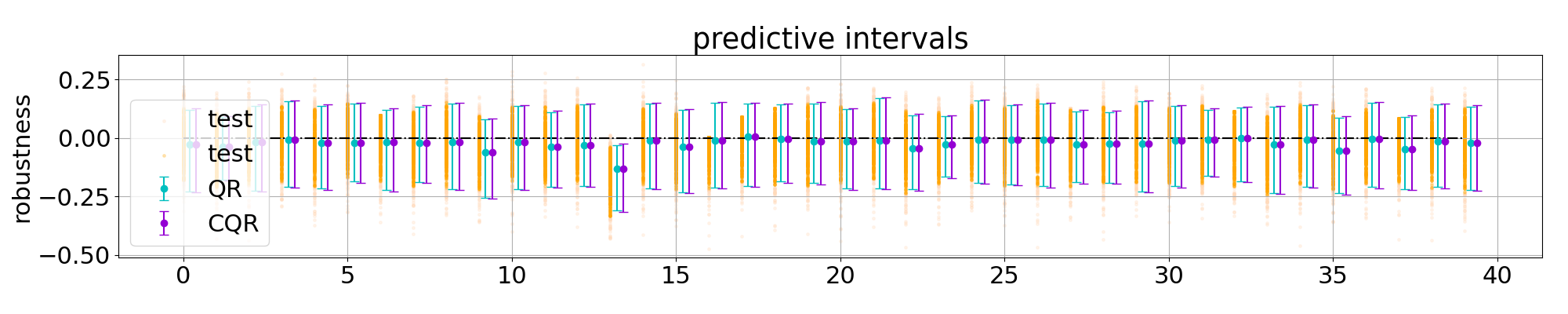}
\includegraphics[width=\textwidth]{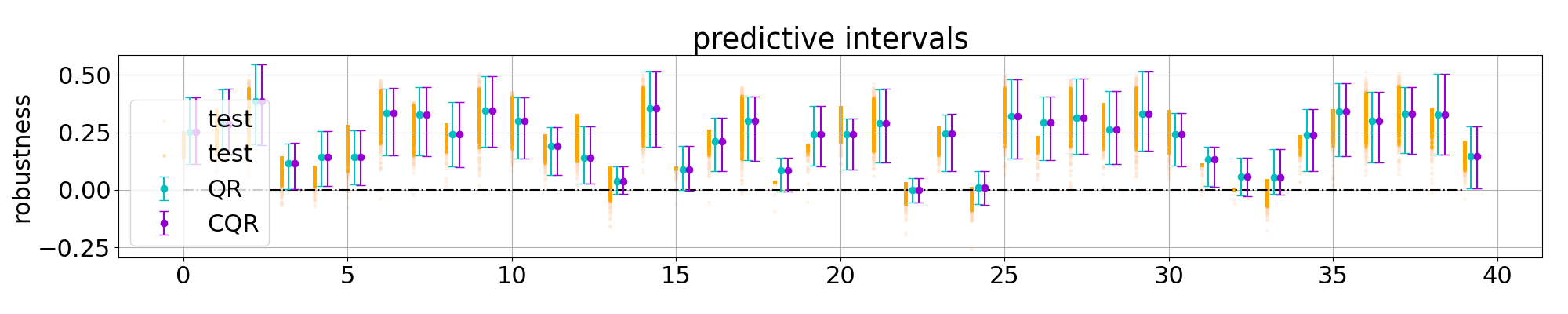}
\caption{Runtime evaluation of the MRH2 model over two randomly selected initial states.
}\label{fig:mrh_seq_plots}
\end{figure*}

\begin{figure*}
\centering

\includegraphics[width=\textwidth]{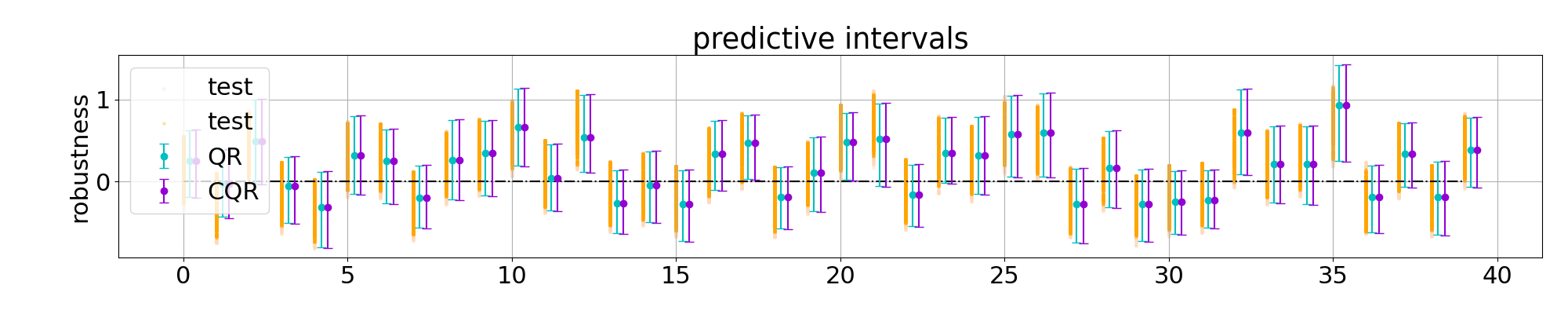}
\includegraphics[width=\textwidth]{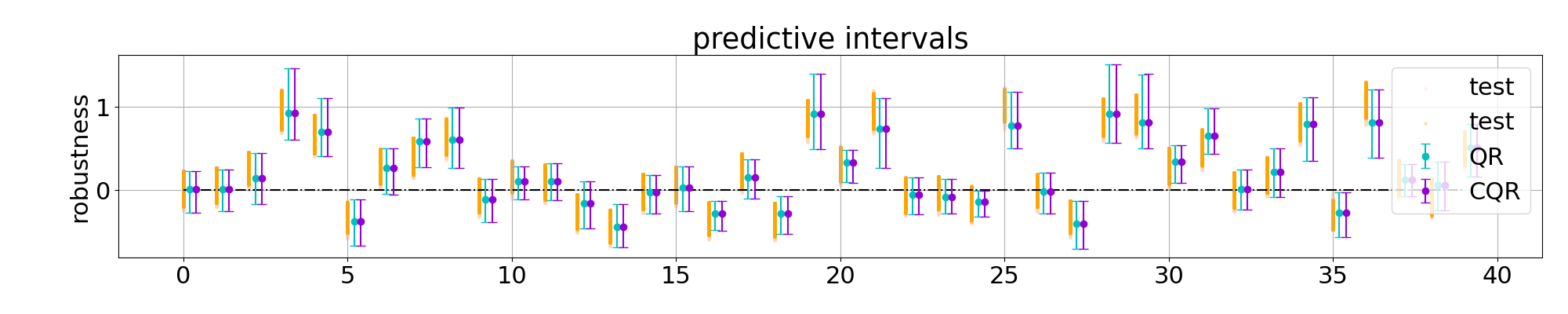}
\caption{Runtime evaluation of the GRN2 model over two randomly selected initial states.
}\label{fig:grn_seq_plots}
\end{figure*}

Moreover, we also show the performances of our QPM at runtime for the following case studies: $AAD$, $MRH2$ and $GRN2$. We randomly sample two trajectories of length $20$ and apply the QPM to each state in the trajectory. Results are visualized in Fig.~\ref{fig:aad_seq_plots}-\ref{fig:grn_seq_plots}.

\end{appendices}

\end{document}